\documentclass[english,aps,pra,a4paper,reprint,floatfix]{revtex4-2}
\usepackage[T1]{fontenc}
\usepackage{float}
\usepackage{placeins}
\usepackage{amsmath}
\usepackage{graphicx}
\usepackage{babel}
\begin{document}
\title{Measurement dependence in a Bell inequality arising from the dynamics
of hidden variables }
\author{Sophia M. Walls and Ian J. Ford}
\email{Correspondence to: sophia.walls.17@ucl.ac.uk}

\affiliation{Department of Physics and Astronomy, University College London, Gower
Street, London, WC1E 6BT, United Kingdom}
\begin{abstract}
Bell inequalities rely on an assumption that the probabilities of
adopting configurations of hidden variables describing a system prior
to measurement are independent of the choice of measured physical
property, also known as measurement independence. Weakening this assumption
could alter the inequalities to accommodate experimental data whilst
maintaining local interactions. A natural avenue for achieving this
would be to model measurement as a dynamical process involving an
interaction between the system and its environment (the measurement
apparatus), that drives the hidden variables towards attractors representing
measurement outcomes of the observable. Implementing such hidden variable
dynamics, we can infer from observed correlations the hidden variable
probability distributions before measurement, which differ according
to which measurement settings were chosen. We explore various models
of the dynamics of the hidden variables under measurement, revealing
features that can create measurement dependence and others that can
not.
\end{abstract}
\maketitle

\section{Introduction}

Bell inequalities arise from analysis of the statistics of purported
`hidden variables' that evolve according to local interactions and
represent `elements of reality' with definite values prior to measurement.
It has been demonstrated that the inequalities can be violated, with
recent experiments removing areas of uncertainty in the analysis such
as the fair-sampling and locality loopholes \citep{Bell64,aspect1982,giustina2015,christensen2013,hensen2015,shalm2015,rowe2001}.
The implication is either that physical effects operate non-locally
between space-like separated points, or that we have to abandon the
concept of a reality independent of observation at microscopic scales
\citep{bell2004,norsen2017}.

Should we wish to avoid these conclusions it is necessary to examine
the assumptions made in Bell's analysis, one of which is `measurement
independence' \citep{hossenfelder2020}, according to which the probabilities
of adopting certain configurations of the hidden variables prior to
measurement are independent of the measurement settings.    We
examine the effect of relaxing this assumption in the standard situation
where two spin half particles in an entangled singlet state have spin
components separately measured along arbitrarily chosen axes $\hat{n}_{1}$
and $\hat{n}_{2}$. We demonstrate, by introducing measurement dependence,
that the upper bound of the Clauser-Horne-Shimony-Holt (CHSH) parameter
may be increased.

The system prior to measurement is specified by a set of hidden variables
$\lambda$ adopted according to a probability density function (pdf)
$\rho(\lambda)$. The assumption of measurement independence is the
use in the analysis of a pdf that depends neither on the measurement
axes $\hat{n}_{1},\hat{n}_{2}$ nor the measurement outcome. This
can seem reasonable but it does not necessarily follow if measurement
is a dynamical process. 

The classical acquisition of information is taken to occur without
changing the system state, while in quantum mechanics the initial
state encodes properties that are extracted with instantaneous projection
of the system to one of the eigenstates of the measured property.
We consider here that measurement could instead involve the nonlinear
evolution of hidden variables towards attractors in their phase space
that correspond to system eigenstates (and device readings correlated
with those eigenstates). The evolution could be deterministic or stochastic.
The point is that dynamics would create a relationship between final
outcomes, represented by an ensemble of post-measurement states described
by hidden variables residing at measurement setting-dependent attractors
in the phase space, and initial states represented by an ensemble
of hidden variable configurations prior to measurement.

The probability distribution of hidden variables before measurement
could therefore differ according to the chosen measurement setting.
In this paper we study the potential for such a mechanism to account
for quantum correlations that break the Bell bound. 

Violation of measurement independence is important since it has the
potential to render vulnerable to attack any entanglement-based technologies
that rely on true quantum randomness \citep{sadhu2023,chaves2021,bierhorst2018,colbeck2012,koh2012,liu2018a,wooltorton2022,horodecki2022,das2021,li2022a,primaatmaja2023,vazirani2014,zapatero2023}.
The effect of measurement dependence on the upper bound of the CHSH
parameter has been explored before \citep{Hall10,putz2014,putz2016,banik2012,supic2022,kim2019a,friedman2019a,hance2022,donadi2022,toner2003,brans1988b,hall2020},
with recent advances demonstrating that violations of the CHSH may
be reproduced without removing all freedom of choice in settings \citep{friedman2019a,banik2012,hall2020}.
Experimental advances have also been made in an attempt to close the
`measurement-dependence loophole' by allowing measurement settings
to be chosen by random or spatio-temporally distant influences \citep{gallicchio2014,handsteiner2017,rauch2018,wu2017,li2018a}.
We nonetheless take the view that the settings, however they are chosen,
can provide information about the system variables prior to measurement,
when the process of measurement is dynamical. The question is then
not \textit{how} the measurement settings are chosen but \textit{how
much} such settings can yield information about the possible distribution
of the system variables prior to measurement. Such information can
then enable an \textit{inference} to be made regarding the probability
distribution of the hidden variables prior to measurement given a
set of chosen measurement settings and knowledge of the distribution
of measurement outcomes. Such inferred distributions are naturally
measurement dependent.  

The plan for the paper is as follows. In section \ref{sec:Inference}
we discuss the process of inference and how it differs from more familiar
processes of probabilistic reasoning. In section \ref{sec:relaxed_CHSH}
we demonstrate that the upper bound of the Clauser-Horne-Shimony-Holt
(CHSH) parameter may raised by relaxing measurement independence.
We consider the usual system of two spin 1/2 particles prepared in
the singlet state where each particle is subjected to measurement
along one of two possible spin axes. The essential ideas are captured
in a toy model described in Section \ref{sec:toy_model}. We consider
measurement outcomes that break the CHSH bound and relate them dynamically
to initial hidden variable conditions. In sections \ref{subsec:analytics}
and \ref{sec:numerics} we illustrate analytically and numerically,
respectively, how the CHSH bound may be raised significantly enough
to account for the quantum correlations. How the upper bound evolves
over time is also explored for a range of different dynamical models.
Some models are able to yield a raised upper bound even when the time
between system preparation and measurement is very long, whilst for
other dynamical models the measurement dependence is destroyed over
time. We explore the conditions for which measurement dependence emerges
and persists long enough to account for quantum correlations.

\section{Backwards in time inference\label{sec:Inference}}

Our reasoning runs backwards in time from information about measurement
outcomes to making inferences about prior probability distributions.
This is perhaps less familiar than relating future outcomes to earlier
situations and a brief discussion is worth having in order not to
confuse the approach with retrocausality.

We can imagine a similar process of inference in a different context.
Consider two couples both of whom had a child some years ago. In the
absence of further information, the likelihood assigned by an observer
to various gender configurations in this situation would naturally
be split equally between four possibilities: girl-girl, boy-boy, boy-girl,
girl-boy. Since the couples are unknown to each other the observer
would not expect the probability distribution over the gender of the
child of one couple to have affected the probability of the gender
of the child of the other. However, if the observer learns that the
couples have a common biological grandchild, this affects the possible
prior probability distributions. A correlation between the genders
of the children emerges with equal likelihoods of girl-boy and boy-girl,
and zero likelihood for the other two. Clearly the existence of a
grandchild common to both couples could not have influenced the probability
distributions for the genders of their children assigned before their
births. The influence we are considering is not retrocausal but rather
a conditioning of past uncertainty given the availability of information
about the present. 

We apply a similar inference process to the Bell experiment of two
entangled spin 1/2 particles, in the singlet state, undergoing measurements
of spin components. If the outcomes of measurements are encoded in
the hidden variables, we may infer the probability distribution of
the hidden variables before measurement, given a set of chosen measurement
settings, an associated distribution of measurement outcomes, and
a model of the dynamics associated with measurement.

These inferred probability distributions do not correspond to the
distributions from which the hidden variable configurations might
have been sampled prior to measurement. Our concern is not with the
selection of such initial configurations. Our reasoning works backward
from measurement outcomes rather than forward from the selection of
initial hidden variables. The distribution reveals to us the necessary
configurations that the hidden variables would have had to adopt in
order to generate the observed quantum correlations according to the
dynamical model. Certain initial configurations of hidden variables
before measurement may be more or less likely to have prevailed, depending
on which measurement axis was chosen and potentially the measurement
outcome. Some hidden variable configurations may even be inaccessible
given a particular choice of measurement and the resulting outcome. 

This reasoning shares commonality with the concept of quantum contextuality,
where depending on the overall measurement context, only a subset
of measurement outcomes that might have been thought possible are
actually realisable \citep{mermin1990,mermin1990a,peres1990}. Measurement
dependence often is attributed to a correlation between the prepared
initial state of the entangled particles and the measurement settings,
such that the initial state somewhat determines which measurement
settings may be chosen \citep{hossenfelder2020,friedman2019a,Hall10,larsson2014}.
We however do not think of measurement dependence in these terms,
but rather that the inferred initial probability distribution of the
hidden variables changes according to the chosen measurement setting,
thus the hidden variables are contextual. The use of non-contextual
hidden variables has previously been identified as a potential loophole
of Bell's theorem \citep{nieuwenhuizen2011,khrennikov2022,kupczynski2020a}.
Since the possible measurement settings are non-commuting, it is natural
that the probability distribution over the hidden variables should
change with each measurement context \citep{nieuwenhuizen2011,khrennikov2022,kupczynski2020a}.
Under this interpretation, measurement dependence can therefore be
regarded as a form of contextuality.

\section{Relaxing the upper bound of the CHSH parameter\label{sec:relaxed_CHSH}}

The CHSH parameter, $S(\hat{a},\hat{b},\hat{a}^{\prime},\hat{b}^{\prime})$,
is defined as \citep{CHSH69}
\begin{equation}
S=\vert C(\hat{a},\hat{b})-C(\hat{a},\hat{b}^{\prime})\vert+\vert C(\hat{a}^{\prime},\hat{b})+C(\hat{a}^{\prime},\hat{b}^{\prime})\vert,\label{eq:1}
\end{equation}
where $C(\hat{a},\hat{b})$ is the correlation function of spin component
measurement outcomes $A$ and $B$ (each taking values $\pm$1) for
particle 1 undergoing a spin measurement along axis $\hat{a}$ and
particle 2 along axis $\hat{b}$, respectively. This may be written
$C(\hat{a},\hat{b})=\overline{AB}=P_{++}^{\hat{a},\hat{b}}-P_{+-}^{\hat{a},\hat{b}}-P_{-+}^{\hat{a},\hat{b}}+P_{--}^{\hat{a},\hat{b}}$,
where $P_{\pm\pm}^{\hat{a},\hat{b}}$ are the probabilities of measurement
outcomes $A=\pm1,B=\pm1$ along the respective axes. Each correlation
function takes a value between $\pm1$ and $S$ could therefore lie
between 0 and 4. 

Assuming that the outcome of measurement of particle 2 is not influenced
by the choice of axis setting nor the outcome of measurement of particle
1, and vice versa, and also assuming that the system prior to measurement
is specified by a set of hidden variables $\lambda$ adopted according
to a probability density function (pdf) $\rho(\lambda)$, then $P_{\pm\pm}^{\hat{a},\hat{b}}=\int p_{1\pm}^{\hat{a}}(\lambda)p_{2\pm}^{\hat{b}}(\lambda)\rho(\lambda)d\lambda$
where $p_{1\pm}^{\hat{a}}(\lambda)$ are the probabilities of outcomes
$\pm1$ for the spin component of particle 1 along axis $\hat{a}$,
for a given specification of hidden variables, and with a similar
meaning for $p_{2\pm}^{\hat{b}}(\lambda)$. 

Measurement independence is the claim that the distribution of the
hidden variables before measurement is not correlated with the chosen
measurement settings: we use a pdf $\rho(\lambda)$ rather than the
conditioned pdf $\rho(\lambda|\hat{n}_{1},\hat{n}_{2})$ where $\hat{n}_{1}$
and $\hat{n}_{2}$ denote axis orientations for measurements on particle
1 and particle 2 respectively. 

Bell's analysis based on these assumptions requires $S$ to have an
upper bound of 2. The crucial observation is that
\begin{equation}
\vert\bar{B}(\hat{b},\lambda)\!-\!\bar{B}(\hat{b}^{\prime},\lambda)\vert+\vert\bar{B}(\hat{b},\lambda)\!+\!\bar{B}(\hat{b}^{\prime},\lambda)\vert\le2,\label{eq:2}
\end{equation}
where $\bar{B}(\hat{b},\lambda)=p_{2+}^{\hat{b}}(\lambda)-p_{2-}^{\hat{b}}(\lambda)$
is the mean outcome of measurement of the spin component of particle
2 along axis $\hat{b}$ for a given set of hidden variables $\lambda$.
Note also that $\vert\bar{B}(\hat{b},\lambda)\vert\le1$ and that
$C(\hat{a},\hat{b})=\overline{AB}=\int\bar{A}(\hat{a},\lambda)\bar{B}(\hat{b},\lambda)\rho(\lambda)d\lambda$.
Since the mean outcome of spin component measurement for particle
1 along axis $\hat{a}$ does not depend on the orientation of axis
$\hat{b}$ specifying the spin component measurement of particle 2,
and vice versa, the average of $AB$ given $\hat{a}$, $\hat{b}$
and $\lambda$ factorises, namely $\overline{AB}(\hat{a},\hat{b},\lambda)=\bar{A}(\hat{a},\lambda)\bar{B}(\hat{b},\lambda)$.

It has nevertheless been shown that the upper bound $S\le2$ can be
violated experimentally, requiring an examination of the assumptions
made in the analysis. We seek to relax the assumption of measurement
independence and demonstrate that this can raise the upper bound of
the CHSH parameter.

The conceptual framework for this presumes that the process of measurement
drives hidden system variables $\lambda$ towards various measurement
setting-dependent attractors in their phase space that correspond
to spin component outcomes for those axis orientations. The chosen
measurement axes and the rules of evolution allow us, in principle,
to deduce the probability density over configurations of $\lambda$
prior to measurement. Since the attractors depend on the measurement
settings, the prior pdf $\rho$ is therefore conditioned on the choice
of axes, such that the correlation function should be written
\begin{equation}
C(\hat{a},\hat{b})=\int\bar{A}(\hat{a},\lambda)\bar{B}(\hat{b},\lambda)\rho(\lambda\vert\hat{a},\hat{b})d\lambda,\label{eq:3}
\end{equation}
using notation to indicate the conditioning of $\rho$. 

The CHSH parameter is built from correlation functions involving four
pairs of measurement axes, and hence four conditioned pdfs over $\lambda$,
which we denote $\rho(\lambda\vert\hat{a},\hat{b})$, $\rho(\lambda\vert\hat{a},\hat{b}^{\prime})$,
$\rho(\lambda\vert\hat{a}^{\prime},\hat{b})$ and $\rho(\lambda\vert\hat{a}^{\prime},\hat{b}^{\prime})$.
We can use these to define a normalised average pdf $\bar{\rho}(\lambda,\hat{a},\hat{b},\hat{a}^{\prime},\hat{b}^{\prime})=\frac{1}{4}\left(\rho(\lambda\vert\hat{a},\hat{b})+\rho(\lambda\vert\hat{a},\hat{b}^{\prime})+\rho(\lambda\vert\hat{a}^{\prime},\hat{b})+\rho(\lambda\vert\hat{a}^{\prime},\hat{b}^{\prime})\right)$
together with combinations that describe the differences between them:
\begin{align}
\bar{\rho}\epsilon & =\frac{1}{4}\!\left(\rho(\lambda\vert\hat{a},\hat{b})-\rho(\lambda\vert\hat{a},\hat{b}^{\prime})+\rho(\lambda\vert\hat{a}^{\prime},\hat{b})-\rho(\lambda\vert\hat{a}^{\prime},\hat{b}^{\prime})\right)\nonumber \\
\bar{\rho}\sigma & =\frac{1}{4}\!\left(\rho(\lambda\vert\hat{a},\hat{b})+\rho(\lambda\vert\hat{a},\hat{b}^{\prime})-\rho(\lambda\vert\hat{a}^{\prime},\hat{b})-\rho(\lambda\vert\hat{a}^{\prime},\hat{b}^{\prime})\right)\nonumber \\
\bar{\rho}\eta & =\frac{1}{4}\!\left(\rho(\lambda\vert\hat{a},\hat{b})-\rho(\lambda\vert\hat{a},\hat{b}^{\prime})-\rho(\lambda\vert\hat{a}^{\prime},\hat{b})+\rho(\lambda\vert\hat{a}^{\prime},\hat{b}^{\prime})\right),\label{eq:7}
\end{align}
such that $\rho(\lambda\vert\hat{a},\hat{b})=\bar{\rho}\left(1+\epsilon+\sigma+\eta\right)$,
$\rho(\lambda\vert\hat{a},\hat{b}^{\prime})=\bar{\rho}\left(1-\epsilon+\sigma-\eta\right)$,
$\rho(\lambda\vert\hat{a}^{\prime},\hat{b})=\bar{\rho}\left(1+\epsilon-\sigma-\eta\right)$,
and $\rho(\lambda\vert\hat{a}^{\prime},\hat{b}^{\prime})=\bar{\rho}\left(1-\epsilon-\sigma+\eta\right)$.
The $\epsilon$, $\sigma$ and $\eta$ functions depend on all four
axis orientations as well as $\lambda$.

Now consider the first combination of correlation functions in the
CHSH parameter, $C(\hat{a},\hat{b})-C(\hat{a},\hat{b}^{\prime})$.
Introducing conditioning of the probability densities according to
the chosen measurement axes this can written as
\begin{align}
 & \int\bar{A}(\hat{a},\lambda)\left(\bar{B}(\hat{b},\lambda)\rho(\lambda\vert\hat{a},\hat{b})-\bar{B}(\hat{b}^{\prime},\lambda)\rho(\lambda\vert\hat{a},\hat{b}^{\prime})\right)d\lambda\nonumber \\
 & =\int\bar{A}(\hat{a},\lambda)\left(\bar{B}(\hat{b},\lambda)-\bar{B}(\hat{b}^{\prime},\lambda)\right)\bar{\rho}(\lambda)d\lambda\nonumber \\
 & +\int\bar{A}(\hat{a},\lambda)\left(\bar{B}(\hat{b},\lambda)+\bar{B}(\hat{b}^{\prime},\lambda)\right)\bar{\rho}(\lambda)\epsilon(\lambda)d\lambda\nonumber \\
 & +\int\bar{A}(\hat{a},\lambda)\left(\bar{B}(\hat{b},\lambda)-\bar{B}(\hat{b}^{\prime},\lambda)\right)\bar{\rho}(\lambda)\sigma(\lambda)d\lambda\nonumber \\
 & +\int\bar{A}(\hat{a},\lambda)\left(\bar{B}(\hat{b},\lambda)+\bar{B}(\hat{b}^{\prime},\lambda)\right)\bar{\rho}(\lambda)\eta(\lambda)d\lambda,\label{eq:8}
\end{align}
which, since $|\bar{A}(\hat{a},\lambda)|\leq1$, implies that
\begin{align}
 & \vert C(\hat{a},\hat{b})-C(\hat{a},\hat{b}^{\prime})\vert\le\int\left|\bar{B}(\hat{b},\lambda)-\bar{B}(\hat{b}^{\prime},\lambda)\right|\bar{\rho}(\lambda)d\lambda\nonumber \\
 & +\int\left|\bar{B}(\hat{b},\lambda)+\bar{B}(\hat{b}^{\prime},\lambda)\right|\bar{\rho}(\lambda)\vert\epsilon(\lambda)\vert d\lambda\nonumber \\
 & +\int\left|\bar{B}(\hat{b},\lambda)-\bar{B}(\hat{b}^{\prime},\lambda)\right|\bar{\rho}(\lambda)\vert\sigma(\lambda)\vert d\lambda\nonumber \\
 & +\int\left|\bar{B}(\hat{b},\lambda)+\bar{B}(\hat{b}^{\prime},\lambda)\right|\bar{\rho}(\lambda)\vert\eta(\lambda)\vert d\lambda.\label{eq:9}
\end{align}
Similarly the combination $C(\hat{a}^{\prime},\hat{b})+C(\hat{a}^{\prime},\hat{b}^{\prime})$
may be written
\begin{align}
 & \int\bar{A}(\hat{a}^{\prime},\lambda)\left(\bar{B}(\hat{b},\lambda)\rho(\lambda\vert\hat{a}^{\prime},\hat{b})+\bar{B}(\hat{b}^{\prime},\lambda)\rho(\lambda\vert\hat{a}^{\prime},\hat{b}^{\prime})\right)d\lambda\nonumber \\
 & =\int\bar{A}(\hat{a}^{\prime},\lambda)\left(\bar{B}(\hat{b},\lambda)+\bar{B}(\hat{b}^{\prime},\lambda)\right)\bar{\rho}(\lambda)d\lambda\nonumber \\
 & +\int\bar{A}(\hat{a}^{\prime},\lambda)\left(\bar{B}(\hat{b},\lambda)-\bar{B}(\hat{b}^{\prime},\lambda)\right)\bar{\rho}(\lambda)\epsilon(\lambda)d\lambda\nonumber \\
 & +\int\bar{A}(\hat{a}^{\prime},\lambda)\left(-\bar{B}(\hat{b},\lambda)-\bar{B}(\hat{b}^{\prime},\lambda)\right)\bar{\rho}(\lambda)\sigma(\lambda)d\lambda\nonumber \\
 & +\int\bar{A}(\hat{a}^{\prime},\lambda)\left(-\bar{B}(\hat{b},\lambda)+\bar{B}(\hat{b}^{\prime},\lambda)\right)\bar{\rho}(\lambda)\eta(\lambda)d\lambda,\label{eq:8-1}
\end{align}
which, since $|\bar{A}(\hat{a},\lambda)|\leq1$, leads to
\begin{align}
 & \vert C(\hat{a}^{\prime},\hat{b})+C(\hat{a}^{\prime},\hat{b}^{\prime})\vert\le\int\left|\bar{B}(\hat{b},\lambda)+\bar{B}(\hat{b}^{\prime},\lambda)\right|\bar{\rho}(\lambda)d\lambda\nonumber \\
 & +\int\left|\bar{B}(\hat{b},\lambda)-\bar{B}(\hat{b}^{\prime},\lambda)\right|\bar{\rho}(\lambda)\vert\epsilon(\lambda)\vert d\lambda\nonumber \\
 & +\int\left|\bar{B}(\hat{b},\lambda)+\bar{B}(\hat{b}^{\prime},\lambda)\right|\bar{\rho}(\lambda)\vert\sigma(\lambda)\vert d\lambda\nonumber \\
 & +\int\left|\bar{B}(\hat{b},\lambda)-\bar{B}(\hat{b}^{\prime},\lambda)\right|\bar{\rho}(\lambda)\vert\eta(\lambda)\vert d\lambda.\label{eq:10}
\end{align}
Combining Eqs. (\ref{eq:9}), (\ref{eq:10}) and (\ref{eq:2}), the
CHSH parameter satisfies
\begin{align}
S & \le2+\mu,\label{eq:12}
\end{align}
where 
\begin{equation}
\mu=2\left(\int\bar{\rho}\left(\vert\epsilon\vert+\vert\sigma\vert+\vert\eta\vert\right)d\lambda\right)\ge0\label{eq:mu}
\end{equation}
represents an elevation of the usual upper limit.  Note that $\epsilon=\sigma=\eta=0$
and hence $\mu=0$ in the absence of measurement dependence. Also,
if only one of the four probability densities is non-zero for a given
$\lambda$, an extreme case of measurement dependence, then $\bar{\rho}\vert\epsilon\vert$,
$\bar{\rho}\vert\sigma\vert$ and $\bar{\rho}\vert\eta\vert$ are
all normalised to unity, in which case $\mu=6$, though it should
be noted that values of $\mu$ above 2 are redundant since $S$ cannot
exceed 4. 

\section{A toy model \label{sec:toy_model}}

We consider two hidden variables $\lambda_{1}$ and $\lambda_{2}$
that separately describe the state of each particle in a Bell experiment.
These variables evolve dynamically under measurement such that trajectories
link the values associated with a measurement outcome to those describing
the initial state of the system at time $-t$. 

\begin{figure}[H]

\centering
\includegraphics[width=1\columnwidth]{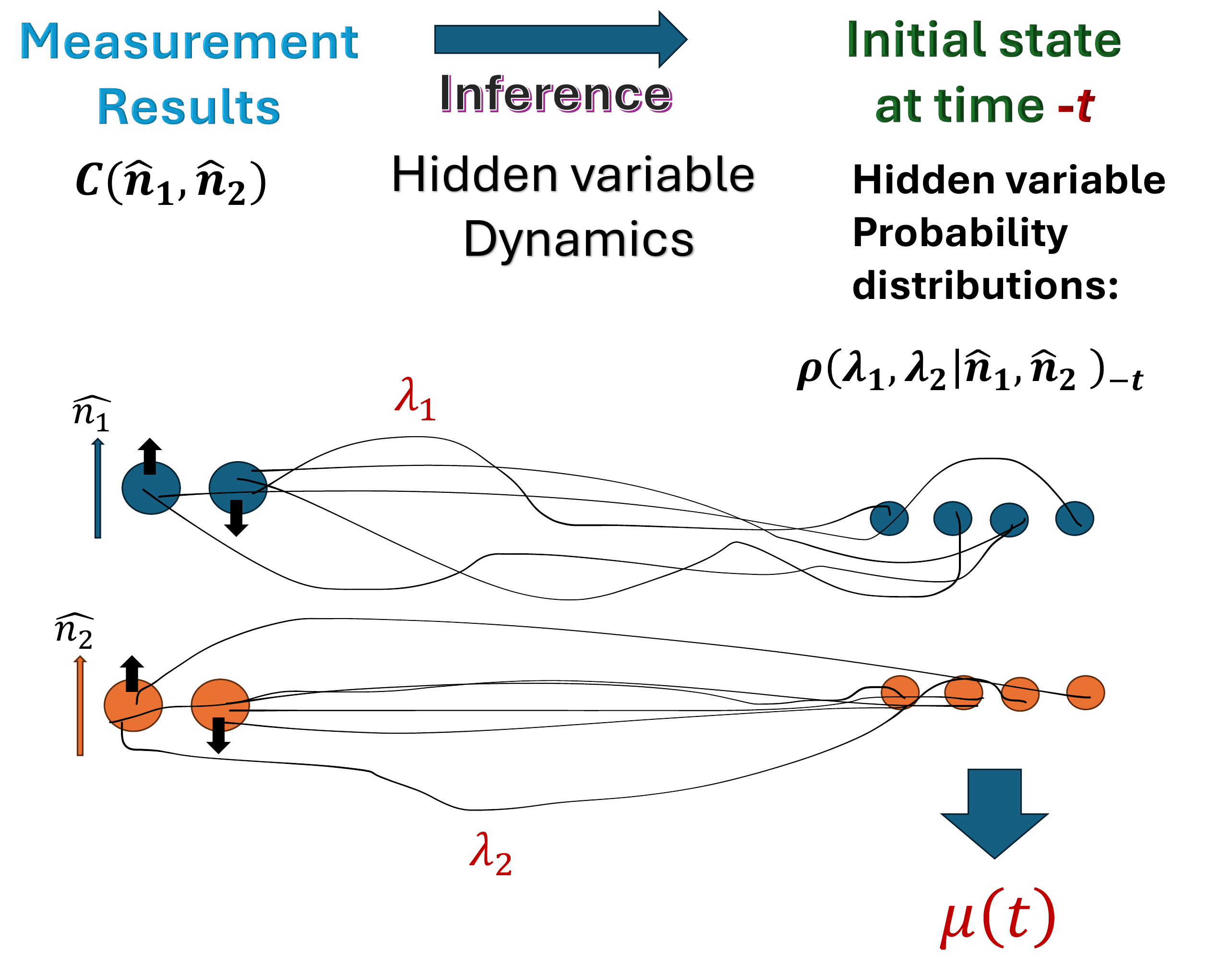}
\caption{The strategy to find the additional term $\mu$ in the upper bound
on the CHSH parameter. We work backwards in time from measurement
outcomes on the left to initial conditions on the right using a model
of the hidden variable dynamics. If the measurement outcomes break
the upper bound on the CHSH parameter, then the initial conditioned
probability distributions for the hidden variables must possess sufficient
measurement setting dependence to raise the bound appropriately. See
text for details. \label{fig:space_time_diag}}

\end{figure}

The post-measurement values of the hidden variables $\lambda_{1}$
and $\lambda_{2}$ are illustrated on the left hand side of Figure
\ref{fig:space_time_diag} as the large blue and orange circles, respectively,
with the associated spin outcome indicated as black arrows. From these
measurement results we work backwards using the hidden variable dynamics
to infer possible initial values of $\lambda_{1}$ and $\lambda_{2}$,
illustrated as the blue and orange circles on the right and connected
to their measurement outcomes by trajectories in the $\lambda_{1}$
and $\lambda_{2}$ phase spaces. We can hence compute the conditioned
probability distributions $\rho(\lambda_{1},\lambda_{2}|\hat{n}_{1},\hat{n}_{2})_{t}$
at the time $-t$ when the system was prepared. From these we are
able to calculate the corresponding additional term $\mu(t)$ in the
CHSH bound using Eq. (\ref{eq:12}). Our aim is to explore whether
the dynamics allows sufficient measurement setting dependence of the
initial probability distributions to raise the bound sufficiently
to accommodate the asserted measurement results.

\begin{figure}[H]
\centering
\includegraphics[width=1\columnwidth]{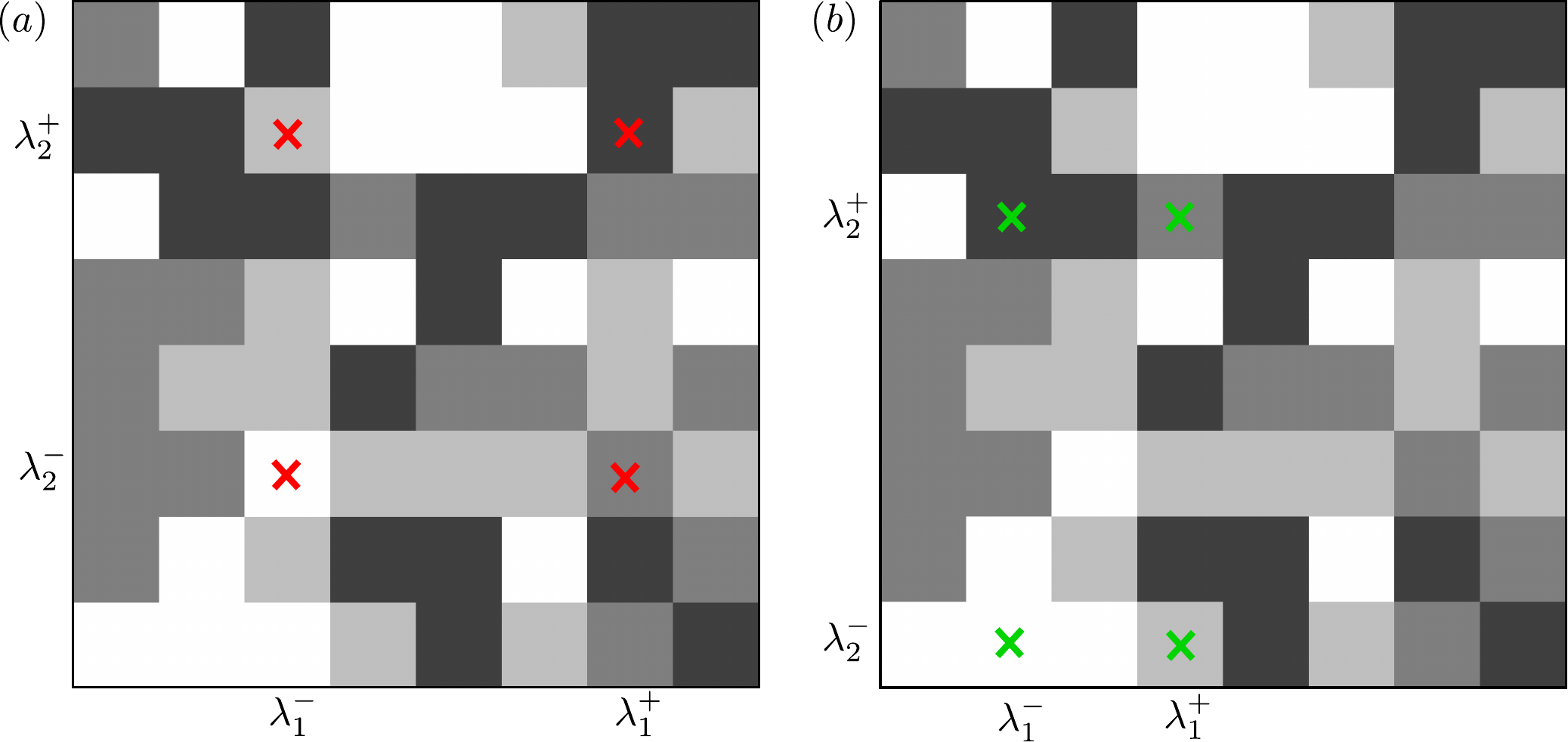}

\caption{A discrete representation of the basins of attraction, over a phase
space of configurations of hidden variables, associated with four
measurement outcomes shown as crosses. All the initial configurations
denoted by one of the four shades of grey evolve under measurement
towards the appropriately shaded attractor coordinate pair $\left(\lambda_{1}^{\pm},\lambda_{2}^{\pm}\right)$
indicated by a cross $\times$. Situations (a) and (b) correspond
to measurement of different system properties with distinct sets (indicated
by colours) of attractors and associated basins. \label{fig:hinterland}}
\end{figure}

To illustrate this situation in a more concrete fashion, we consider
two discrete hidden variables $\lambda_{1}$ and $\lambda_{2}$ and
a phase space in the form of a grid of squares labelled with integer
values. $\lambda_{1}$ describes the state of the first particle in
the entangled spin 1/2 particle pair and $\lambda_{2}$ the second.
The measurement of properties $A$ and $B$ of particle 1 and particle
2 respectively, is represented by attraction under the dynamics of
each of $\lambda_{1}$ and $\lambda_{2}$ towards one of two points,
yielding four `targets' at coordinates $\left(\lambda_{1}^{\pm},\lambda_{2}^{\pm}\right)$.
These represent the four possible combinations of spin measurement
outcomes of the entangled particle pair: the situation is illustrated
in Figure \ref{fig:hinterland}. The idea is that interactions between
the system and measuring device bring about a dynamical evolution
of the $\lambda_{1}$ and $\lambda_{2}$ to attractors (here fixed
points) located at the targets. The outcome of the measurement of
each property is $\pm$1, as designated by the superscripts on the
target coordinates. 

We consider a situation where passage to each target arises from separate
basins of attraction in the phase space \citep{palmer1997}. These
are shown in Figure \ref{fig:hinterland}(a) as four shades of grey:
the measurement dynamics can generate trajectories starting from any
of the black squares, for example, all of which terminate at the top
right target $(\lambda_{1}^{+},\lambda_{2}^{+})$.

The probabilities of adopting each black square prior to measurement
would accumulate under the dynamics to generate a probability $P_{++}$
of an outcome $A=B=+1$ associated with arrival of the system at the
$(\lambda_{1}^{+},\lambda_{2}^{+})$ target, post-measurement. The
suffixes in this probability correspond to the superscripts on the
target coordinates. Conversely, if we wish to deduce the probabilities
of having started out on a black square conditioned on information
about which measurement is performed and its outcome, then we need
to distribute the probability $P_{++}$ of arrival at $(\lambda_{1}^{+},\lambda_{2}^{+})$
after measurement over the basin of attraction of that target at earlier
times, again in accordance with the measurement dynamics.

We can now see how such conditioned probabilities depend on the measurement
concerned, represented in the toy model as attraction towards a different
set of targets. Consider targets represented by the green crosses
in Figure \ref{fig:hinterland}(b). The basins of attraction under
the measurement dynamics might form a different pattern for these
targets. There would be different probabilities of adoption of configurations
prior to measurement when conditioned on a measurement defined by
the green crosses. This is compounded if we consider probabilities
of arrival $P_{\pm\pm}$ (probabilities associated with measurement
outcomes) at the green and red targets that differ from one another. 

It is crucial for the emergence of measurement dependence in this
toy model that the phase space should be separated into basins of
attraction to each outcome target for each measurement situation.
We shall investigate cases where this is not the case later on and
find that there is then insufficient measurement dependence in the
initial probability distributions to accommodate the asserted measurement
outcomes. In systems where the evolution is deterministic, outcomes
are encoded in the coordinates describing the initial state and basins
of attraction would be natural. For systems governed by stochastic
dynamics such encoding could persist to some degree depending on the
situation. 

\begin{figure}[H]
\centering
\includegraphics[width=1\columnwidth]{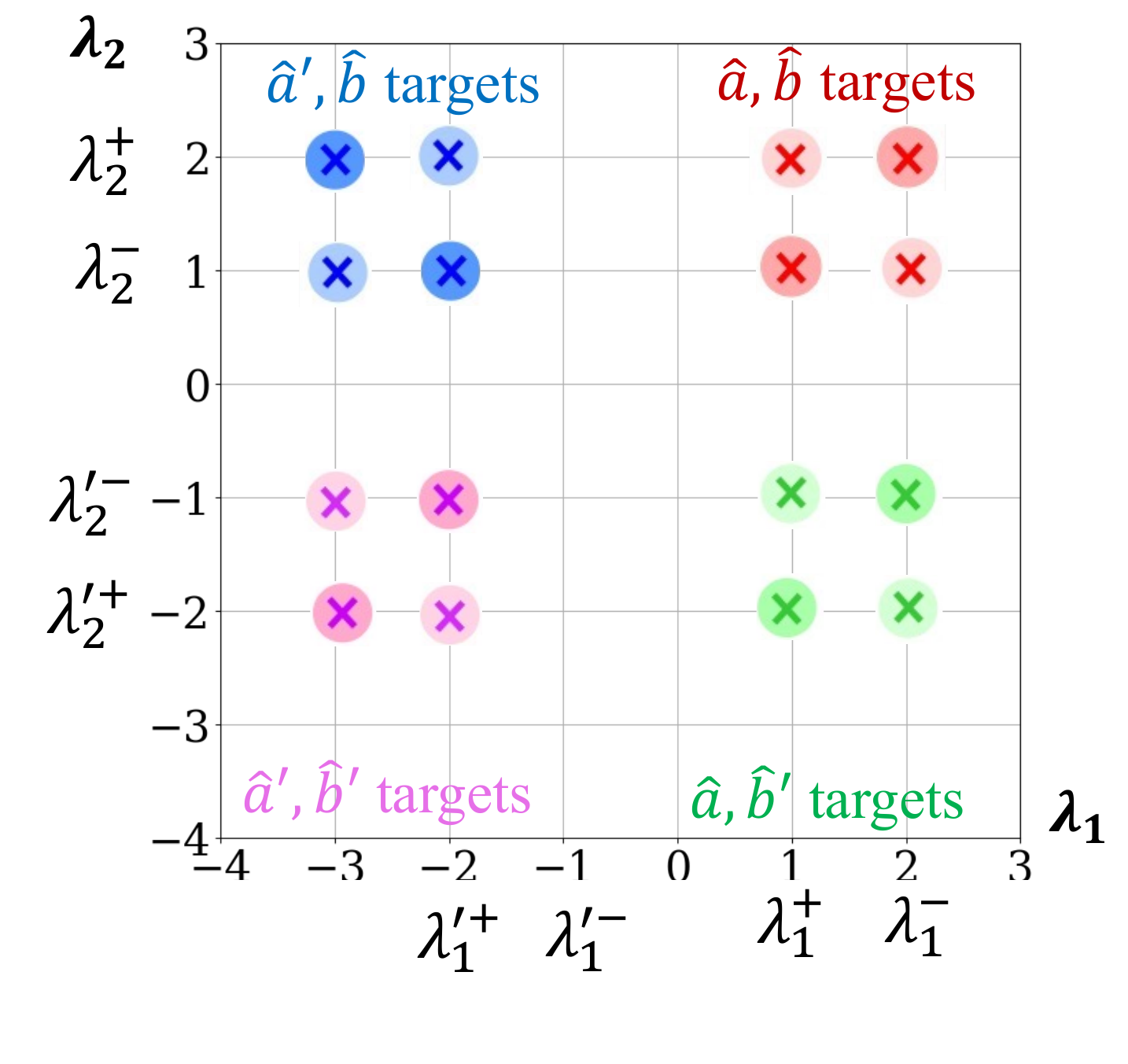}

\caption{Configurations of the toy model are represented by grid squares in
a two dimensional discrete phase space with integer coordinates $\lambda_{1}$
and $\lambda_{2}$. The process of measurement draws the configuration
towards one of four targets, shown as crosses in a specific colour,
representing the four outcomes of measurement of properties $A$ and
$B$. Four different experimental circumstances are designated by
labels $\hat{a},\hat{b}$, etc, and distinguished by different colours.
Arrival at one of the targets yields an outcome $A=\pm1$ and $B=\pm1$
according to the superscripts of the target coordinates. The probability
of such an outcome is indicated by coloured circles at the targets:
the darker the shade, the higher the probability $P_{\pm\pm}^{\hat{x},\hat{y}}$
of arrival at that target. \label{fig:Four-targets}}
\end{figure}

We now consider four possible measurement situations: settings $\hat{a},\hat{a}'$
with outcomes $A$ concern particle 1, and $\hat{b},\hat{b}'$ with
outcomes $B$ concern particle 2. We illustrate the possible measurement
setting combinations and their associated distributions of measurement
outcomes in Figure \ref{fig:Four-targets}.  We define four sets of
targets for the measurement dynamics that evolve the variables $\lambda_{1}$
and $\lambda_{2}$. The target pairs $\lambda_{1}^{\pm}$ and $\lambda_{1}^{\prime\pm}$
will be associated with `axis choices' $\hat{a}$ and $\hat{a}^{\prime}$,
respectively, and $\lambda_{2}^{\pm}$ and $\lambda_{2}^{\prime\pm}$
are to be associated with `axes' $\hat{b}$ and $\hat{b}^{\prime}$.
The terminology is chosen to establish an analogy with spin component
measurement, and the superscripts indicate the outcome values $\pm1$
for properties $A$ and $B$. The four measurement situations are
specified by the choice of blue, red, pink or green targets in Figure
\ref{fig:Four-targets}. We associate +1 outcomes with targets situated
at even label positions on the grid and $-1$ outcomes with odd labels. 

We can then consider a set of probabilities $P_{\pm\pm}^{\hat{x},\hat{y}}$
for post-measurement arrival of the system at each of the four targets
in each group, for measurement settings $\hat{x}=\hat{a}$ or $\hat{a}^{\prime}$
and $\hat{y}=\hat{b}$ or $\hat{b}^{\prime}$. These represent the
probability distributions of measurement outcomes for a given pair
of measurement settings. Figure \ref{fig:Four-targets} shows the
targets as circular symbols with different shades of colour to indicate
the magnitude of the probability of arrival at that target. To simplify
the situation, we impose conditions $P_{++}^{\hat{x},\hat{y}}=P_{--}^{\hat{x},\hat{y}}$
and $P_{+-}^{\hat{x},\hat{y}}=P_{-+}^{\hat{x},\hat{y}}$. Furthermore,
we consider identical outcome probabilities for the red, blue and
pink measurement situations, namely $P_{\pm\pm}^{\hat{a},\hat{b}}=P_{\pm\pm}^{\hat{a}^{\prime},\hat{b}}=P_{\pm\pm}^{\hat{a}^{\prime},\hat{b}^{\prime}}$.
However, we design the green measurement situation to be different
by setting $P_{\pm\mp}^{\hat{a},\hat{b}^{\prime}}=P_{\pm\pm}^{\hat{a},\hat{b}}$,
namely that dark and light shades of green characterise measurement
outcomes coloured in light and dark shades, respectively, for the
other three cases. This is to enable values of the CHSH parameter
that exceed a value of two.

We can combine this description of measurement outcomes with a chosen
pattern of basins of attraction associated with each target to derive
the conditioned probabilities over the $(\lambda_{1},\lambda_{2})$
phase space prior to measurement for each pair of chosen `axes'. Our
aim is to produce a probability distribution $P(\lambda_{1},\lambda_{2}|\hat{a},\hat{b}^{\prime})_{-t}$
over the phase space at time $-t$ that is conditioned on outcomes
at the green targets, and which differs from those conditioned on
the blue, red and purple targets, which for simplicity we shall choose
to be the same, namely $P(\lambda_{1},\lambda_{2}|\hat{a},\hat{b})_{-t}=P(\lambda_{1},\lambda_{2}|\hat{a}^{\prime},\hat{b})_{-t}=P(\lambda_{1},\lambda_{2}|\hat{a}^{\prime},\hat{b}^{\prime})_{-t}\neq P(\lambda_{1},\lambda_{2}|\hat{a},\hat{b}^{\prime})_{-t}$.
The prior distribution over the hidden variables $\lambda_{1}$ and
$\lambda_{2}$ then clearly depends on the measurement situation.
Specifically, for the green measurement situation, there is a higher
probability associated with the anti-correlated measurement outcomes,
whereas correlated outcomes would be more probable in the other three
measurement situations.

We then investigate whether the CHSH parameter implied by the toy
model is compatible with a raised upper limit arising from the inferred
measurement dependence and the additional term $\mu$ in Eq. (\ref{eq:12}). 

In order to represent the locality of measurement dynamics we require
that the hidden variables should evolve independently of one another.
The marginal distribution $P(\lambda_{1}|\hat{n_{1}})_{-t}=\sum_{\lambda_{2}}P(\lambda_{1},\lambda_{2}|\hat{n_{1}},\hat{n}_{2})_{-t}$
is therefore independent of the measurement axis chosen for the other
particle (and similarly for $P(\lambda_{2}|\hat{n}_{2})_{-t}$). 

Finally, we reiterate that the ensembles of initial configurations
of the system are inferred distributions of hidden variables conditioned
on the measurement settings and dynamics presumed to operate. This
should not be confused with a situation where nature somehow prepares
the system with awareness of the yet to be made choice of the measurement
settings. This is not what we are considering. The earlier discussion
of the probability distributions of the genders of children borne
to couples conditioned or not conditioned on whether they possess
a common grandchild, illustrates this point. A grandchild does not
influence the probabilities that governed the gender of its parents
before their birth. The probability distributions in question simply
represent the availability of relevant information at a later time.

We actually have no interest in probability distributions of hidden
variables that are \emph{not} conditioned on the measurement settings.
It is the conditioned distributions that are relevant to the various
measurement setting dependent correlation functions used in building
the CHSH parameter.

\section{Asymptotic value of $\mu$ for symmetric random walk dynamics in
the toy model \label{subsec:analytics}}

Let us first consider a measurement dynamics consisting of independent,
symmetric random walks of the two hidden variables in two dimensions
with equal probabilities of moving left or right with step-size $\vert\Delta\lambda_{1}\vert=\vert\Delta\lambda_{2}\vert=1$
per time-step, terminating at a target. The scheme means that working
backwards for an even number of time-steps, the possible values of
$\lambda_{1}$ and $\lambda_{2}$ (the basin of attraction) leading
to a target with even values of $\lambda_{1}$ and $\lambda_{2}$
(an even-even target) also take even values, whereas for an odd number
of time-steps $\lambda_{1}$ and $\lambda_{2}$ would both adopt odd
values. Similarly, the possible values of $\lambda_{1}$ and $\lambda_{2}$
leading to an even-odd target will be even and odd respectively for
an even number of time-steps and odd and even respectively for an
odd number of time-steps. The even-oddness (parity) of possible values
of $\lambda_{1}$ and $\lambda_{2}$ leading to a target therefore
alternates with the time-step and from hereon we shall consider the
situation at an even time-step. Thus we have a dynamics that defines
certain basins of attraction for given outcomes on targets associated
with specified measurement settings. For simplicity, we consider measurement
to take place over a very large but even number of steps, namely $t\to\infty$.
To this end we drop the suffix $-t$ on the conditioned prior probability
distribution over the phase space.

We also assume a phase space with an even number of locations in each
dimension ($N$ locations in all) and periodic boundary conditions.
This means that the probability $P(\lambda_{1,2}|\hat{a},\hat{b})$
at any even-even location in the phase space, prior to measurement
with settings $\hat{a},\hat{b}$, is a constant $P_{ee}^{\hat{a},\hat{b}}$
given by the probability of arrival $P_{++}^{\hat{a},\hat{b}}$ at
red target $(\lambda_{1}^{+},\lambda_{2}^{+})$ divided by the number
of such locations, $N/4$, namely $P_{ee}^{\hat{a},\hat{b}}=4P_{++}^{\hat{a},\hat{b}}/N$.
Similarly, the conditioned probability $P_{eo}^{\hat{a},\hat{b}}$
at an even-odd location on the grid prior to measurement is the arrival
probability $P_{+-}^{\hat{a},\hat{b}}$ at red target $(\lambda_{1}^{+},\lambda_{2}^{-})$
divided equally between all even-odd locations in the grid, or $P_{eo}^{\hat{a},\hat{b}}=4P_{+-}^{\hat{a},\hat{b}}/N$.
The conditioned probability at an odd-even location is $P_{oe}^{\hat{a},\hat{b}}=4P_{-+}^{\hat{a},\hat{b}}/N$
and for odd-odd locations it is $P_{oo}^{\hat{a},\hat{b}}=4P_{--}^{\hat{a},\hat{b}}/N$.
The same arguments apply to measurement situation $\hat{a},\hat{b}^{\prime}$
with arrival probabilities $P_{\pm\pm}^{\hat{a},\hat{b}^{\prime}}$
at green targets, measurement situation $\hat{a}^{\prime},\hat{b}$
with arrival probabilities $P_{\pm\pm}^{\hat{a}^{\prime},\hat{b}}$
at blue targets, and measurement situation $\hat{a}^{\prime},\hat{b}^{\prime}$
with arrival probabilities $P_{\pm\pm}^{\hat{a}^{\prime},\hat{b}^{\prime}}$
at purple targets.

\begin{figure}[H]
\centering
\includegraphics[width=1\columnwidth]{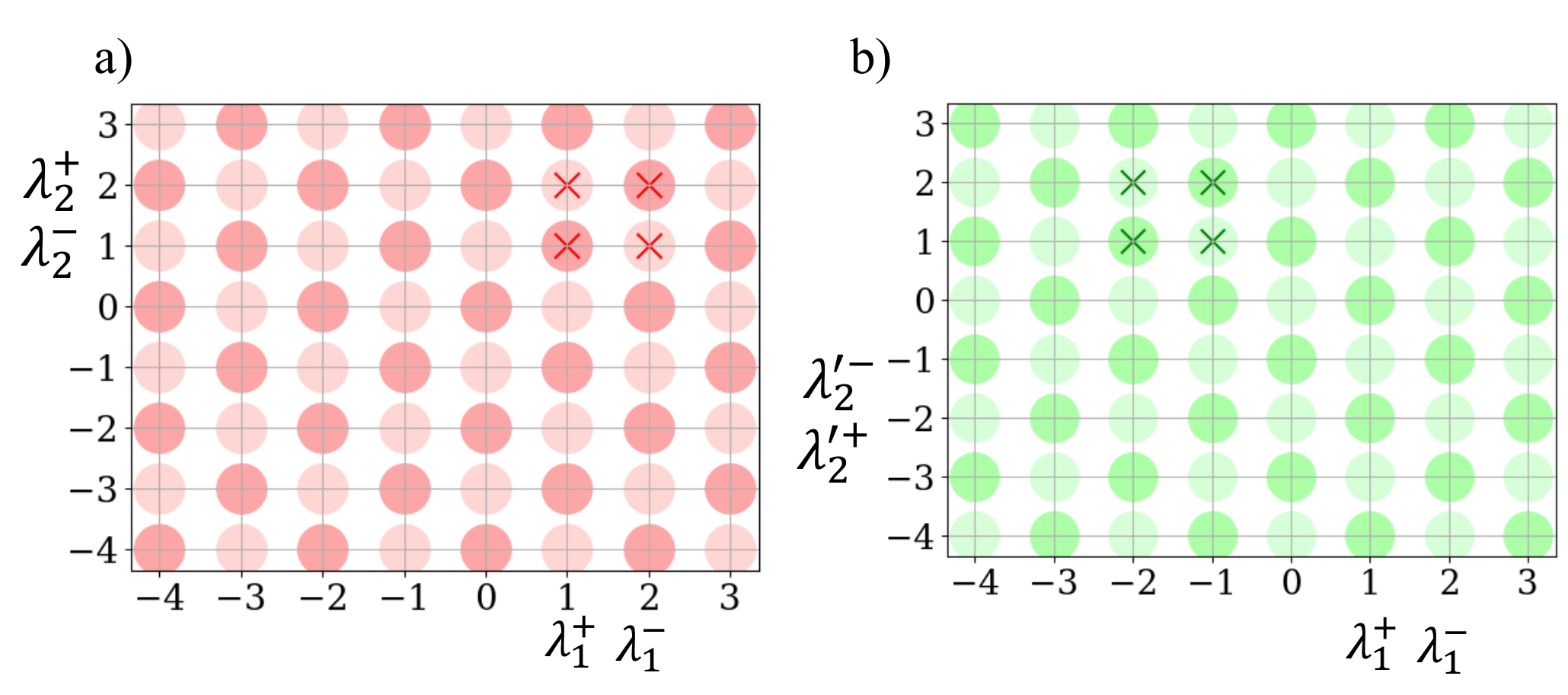}

\caption{Probability distributions characterising the adoption of hidden variable
configurations for a very large even number of timesteps prior to
measurement of property $AB$ in experimental circumstances (a) $\hat{a},\hat{b}$
and (b) $\hat{a},\hat{b}^{\prime}$, shown on the left and right in
red and green, respectively, with the shades representing magnitude.
These are conditioned on (i) the locations of the targets $\lambda_{i}^{\pm}$,
$\lambda_{i}^{\prime\pm}$ in the grid to which the configurations
are attracted by measurement dynamics consisting of independent symmetric
random walks in $\lambda_{1}$ and $\lambda_{2}$ with a step length
of 1, and (ii) on the overall probabilities of reaching each of the
targets. These are asymptotic distributions of probability after taking
an even number of backward time-steps. Notice that the two conditioned
probability distributions differ, which is measurement dependence.
\label{fig:red-and-green}}
\end{figure}

What this means is that the conditioned probabilities on the phase
space are repeated versions of the patterns of probabilities of arrival
at the targets after measurement. This is illustrated in Figure \ref{fig:red-and-green}(a)
for the $\hat{a},\hat{b}$ measurement situation. The dark and light
shades of red represent the two probabilities $P_{ee}^{\hat{a},\hat{b}}$
and $P_{eo}^{\hat{a},\hat{b}}$, respectively, recalling that we have
imposed $P_{ee}^{\hat{a},\hat{b}}=P_{oo}^{\hat{a},\hat{b}}$ and $P_{eo}^{\hat{a},\hat{b}}=P_{oe}^{\hat{a},\hat{b}}$.
The distribution $P(\lambda_{1},\lambda_{2}|\hat{a},\hat{b})$ illustrated
is the discrete analogue of the pdf $\rho(\lambda\vert\hat{a},\hat{b})$
considered for the situation with a continuum of hidden variable values.
Under our assumptions, the conditioned initial probability distributions
for measurement situations $\hat{a}^{\prime},\hat{b}$ and $\hat{a}^{\prime},\hat{b}^{\prime}$
are identical to the distribution shown for $\hat{a},\hat{b}$ (but
would be illustrated in blue and pink, respectively). However, the
conditioned probability distribution for the situation $\hat{a},\hat{b}^{\prime}$
shown in Figure \ref{fig:red-and-green}(b) is different. A light
shade of green, representing $P_{ee}^{\hat{a},\hat{b}^{\prime}}=P_{eo}^{\hat{a},\hat{b}}$,
lies at a location where a dark shade of red, representing $P_{ee}^{\hat{a},\hat{b}}$,
is found in Figure \ref{fig:red-and-green}(a), and vice versa. 

The analogue of the additional term in Eq. (\ref{eq:12}) for the
toy model is a sum of magnitudes of differences between the probability
distribution shown in red in Figure \ref{fig:red-and-green}(a) and
its blue, pink and green counterparts, analogous to $\rho(\lambda\vert\hat{a},\hat{b})$,
$\rho(\lambda\vert\hat{a}^{\prime},\hat{b})$, $\rho(\lambda\vert\hat{a}^{\prime},\hat{b}^{\prime})$
and $\rho(\lambda\vert\hat{a},\hat{b}^{\prime})$, respectively. For
the simplified case under consideration, the first three distributions
are identical and the fourth is different. This implies, in the notation
of a continuum phase space, that $\bar{\rho}\epsilon=-\bar{\rho}\sigma=\bar{\rho}\eta=\frac{1}{4}\left(\rho(\lambda\vert\hat{a},\hat{b})-\rho(\lambda\vert\hat{a},\hat{b}^{\prime})\right)$
and hence the additional term in the Bell inequality for $S$ in these
circumstances is $\mu=\frac{3}{2}\int\vert\rho(\lambda\vert\hat{a},\hat{b})-\rho(\lambda\vert\hat{a},\hat{b}^{\prime})\vert d\lambda$.

For the discrete phase space the integral is replaced by a sum of
moduli of the differences in probability at each position of the grid
conditioned on the red and green measurement outcomes. For the situation
we have constructed, these differences are $P_{ee}^{\hat{a},\hat{b}}-P_{ee}^{\hat{a},\hat{b}^{\prime}}$
at even-even and odd-odd points on the grid, and $P_{eo}^{\hat{a},\hat{b}}-P_{eo}^{\hat{a},\hat{b}^{\prime}}$
at even-odd and odd-even points, and these quantities are given by
$\pm\left(P_{ee}^{\hat{a},\hat{b}}-P_{eo}^{\hat{a},\hat{b}}\right)$,
respectively. The sum of moduli of probability differences over the
grid is therefore $N\left|P_{ee}^{\hat{a},\hat{b}}-P_{eo}^{\hat{a},\hat{b}}\right|=4\left|P_{++}^{\hat{a},\hat{b}}-P_{+-}^{\hat{a},\hat{b}}\right|$
and the additional term for the toy model with the dynamics specified
is $\mu=6\left|P_{++}^{\hat{a},\hat{b}}-P_{+-}^{\hat{a},\hat{b}}\right|$.

Similarly, all four correlation functions can be specified in terms
of $P_{++}^{\hat{a},\hat{b}}$ and $P_{+-}^{\hat{a},\hat{b}}$, specifically
$C(\hat{a},\hat{b})=C(\hat{a}^{\prime},\hat{b})=C(\hat{a}^{\prime},\hat{b}^{\prime})=-C(\hat{a},\hat{b}^{\prime})=2\left(P_{++}^{\hat{a},\hat{b}}-P_{+-}^{\hat{a},\hat{b}}\right)$.
Thus the CHSH parameter is $S=4\vert C(\hat{a},\hat{b})\vert$ and
the additional term is $\mu=3\vert C(\hat{a},\hat{b})\vert$. Bell's
analysis therefore requires that
\begin{equation}
S=4\vert C(\hat{a},\hat{b})\vert\le2+\mu=2+3\vert C(\hat{a},\hat{b})\vert,\label{eq:13}
\end{equation}
which is clearly satisfied for the relevant range $0\le\vert C(\hat{a},\hat{b})\vert\le1$.
We conclude that measurement dependence created by the dynamics of
the hidden variables during the measurement process has elevated the
upper limit to accommodate any imposed value of the CHSH parameter.
Furthermore, the conditioning of the hidden variables on the measurement
settings is retained even at an asymptotic time limit, which is due
to the existence of basins of attraction.

\section{Numerical realisations of toy model dynamics \label{sec:numerics}}

We now calculate the value of the upper bound of the additional term
numerically for different models of the measurement dynamics and for
non-asymptotic measurement conditions. We consider probability distributions
over the hidden variables $P(\lambda_{1},\lambda_{2}|\hat{n}_{1},\hat{n}_{2})_{-t}$
at a time $-t$ relative to the measurement at $t=0$ for measurement
settings $\hat{n}_{1}$ and $\hat{n}_{2}$ for the first and second
particle, respectively, writing 

\begin{align}
P(\lambda_{1},\lambda_{2}|\hat{n}_{1},\hat{n}_{2})_{-t} & =P_{++}^{\hat{n}_{1},\hat{n}_{2}}p(\lambda_{1}|\lambda_{1}^{+},\hat{n}_{1})_{-t}p(\lambda_{2}|\lambda_{2}^{+},\hat{n}_{2})_{-t}\nonumber \\
+ & P_{-+}^{\hat{n}_{1},\hat{n}_{2}}p(\lambda_{1}|\lambda_{1}^{-},\hat{n}_{1})_{-t}p(\lambda_{2}|\lambda_{2}^{+},\hat{n}_{2})_{-t}\nonumber \\
+ & P_{+-}^{\hat{n}_{1},\hat{n}_{2}}p(\lambda_{1}|\lambda_{1}^{+},\hat{n_{1}})_{-t}p(\lambda_{2}|\lambda_{2}^{-},\hat{n}_{2})_{-t}\nonumber \\
+ & P_{--}^{\hat{n}_{1},\hat{n}_{2}}p(\lambda_{1}|\lambda_{1}^{-},\hat{n}_{1})_{-t}p(\lambda_{2}|\lambda_{2}^{-},\hat{n}_{2})_{-t},\label{eq:calculate pdfs}
\end{align}
where $p(\lambda_{i}|\lambda_{i}^{\pm},\hat{n}_{i})_{-t}$ is the
conditional probability distribution over $\lambda_{i}$ at time $-t$
generated by the backward dynamics of $\lambda_{i}$ given that at
$t=0$ the system resides at target value $\lambda_{i}^{\pm}$ appropriate
to setting $\hat{n}_{i}$. As before, the $P_{\pm\pm}^{\hat{n}_{1},\hat{n}_{2}}$
are the outcome probabilities for measurement settings $\hat{n}_{1}$
and $\hat{n}_{2}$, as illustrated in Figure \ref{fig:Four-targets}.

We impose periodic boundaries on the values of $(\lambda_{1},\lambda_{2})$
in such a way that they maintain their parity: even-even values outside
the boundary get mapped to even-even values within the boundary and
similarly for odd-odd, even-odd and odd-even coordinates. We set the
size of the $(\lambda_{1},\lambda_{2})$ phase space to be $8\times8$,
with $\lambda_{1,2}$ taking values between $-4$ and $+3$.  This
can be achieved using $\lambda_{PB}=((\lambda+4)\oplus8)-4$ where
$\lambda_{PB}$ is the value of $\lambda$ after mapping through the
periodic boundaries and $\oplus$ represents a modulo function.

We present three models of measurement in the toy model that illustrate
some of the issues. Further cases are studied in Appendices \ref{sec:Symmetric-random-walk}
and \ref{sec:Asymmetric-random-walk}.

\subsection{Symmetric random walk for finite time}

We first calculate the probability distributions $p(\lambda_{1}|\lambda_{1}^{\pm},\hat{n}_{1})_{-t}$
and $p(\lambda_{2}|\lambda_{2}^{\pm},\hat{n}_{2})_{-t}$ for symmetric
random walks with step size $|\Delta\lambda|=1$ per time-step, namely
the model considered in section \ref{subsec:analytics} but now for
finite time. The probability of adopting a certain value of $\lambda_{i}$
at a time $-t$ corresponding to an even number of backward time-steps,
given that the system reaches a target $\lambda_{i}^{\pm}$ at $t=0$
for a chosen measurement axis $\hat{n}$, is then 

\begin{equation}
p(\lambda_{i}|\lambda_{i}^{\pm},\hat{n})_{-t}=\frac{1}{2^{t}}\frac{t!}{\left(\frac{t-\lambda_{i}+\lambda_{i}^{\pm}}{2}\right)!\left(\frac{t+\lambda_{i}-\lambda_{i}^{\pm}}{2}\right)!}\label{eq:symmetric_walk}
\end{equation}

The possible starting points are then $\lambda_{i}^{\pm}-t,\lambda_{i}^{\pm}-t+2,\lambda_{i}^{\pm}-t+4,....,\lambda_{i}^{\pm}+t-4,\lambda_{i}^{\pm}+t-2,\lambda_{i}^{\pm}-t$.
The probability distribution over both $\lambda_{1}$ and $\lambda_{2}$
can then be found through Eq. (\ref{eq:calculate pdfs}) using the
probabilities of arrival given in Figure \ref{fig:Four-targets} and
considering each of the possible targets (measurement outcomes) for
a given measurement situation. 

\begin{figure}[H]
\centering
\includegraphics[width=1\columnwidth]{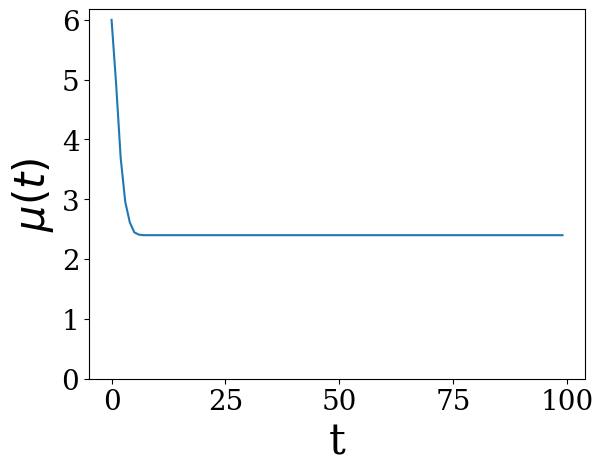}

\caption{The numerically calculated additional term $\mu(t)$ against time
$t$ running backwards from the moment of measurement at $t=0$. The
dynamics consist of independent symmetric random walks in $\lambda_{1}$
and $\lambda_{2}$ with a step size of unity. We use post-measurement
outcome probabilities $P_{\pm\pm}^{\hat{a},\hat{b}}=P_{\pm\pm}^{\hat{a}',\hat{b}}=P_{\pm\pm}^{\hat{a}',\hat{b}'}=P_{\pm\mp}^{\hat{a},\hat{b}'}=0.45$
and $P_{\pm\mp}^{\hat{a},\hat{b}}=P_{\pm\mp}^{\hat{a}',\hat{b}}=P_{\pm\mp}^{\hat{a}',\hat{b}'}=P_{\pm\pm}^{\hat{a},\hat{b}'}=0.05$.\label{fig:sym_add_term_time}}

\end{figure}

We calculate the additional term $\mu$ in the Bell limit arising
from the measurement dependence in the probability distributions $P(\lambda_{1},\lambda_{2}|\hat{n}_{1},\hat{n}_{2})_{-t}$
given by Eq. (\ref{eq:7}). Figure \ref{fig:sym_add_term_time} shows
that the additional term decreases to a value of around $\approx2.4$,
where it remains, even at $t=100$, revealing that the conditioning
of the probabilities of adopting hidden variables prior to measurement
is persistent for this dynamical model, as was expected from the analytical
calculations in section \ref{subsec:analytics}. 

\begin{figure}[H]
\centering
\textbf{A)}\\
\includegraphics[width=1\columnwidth]{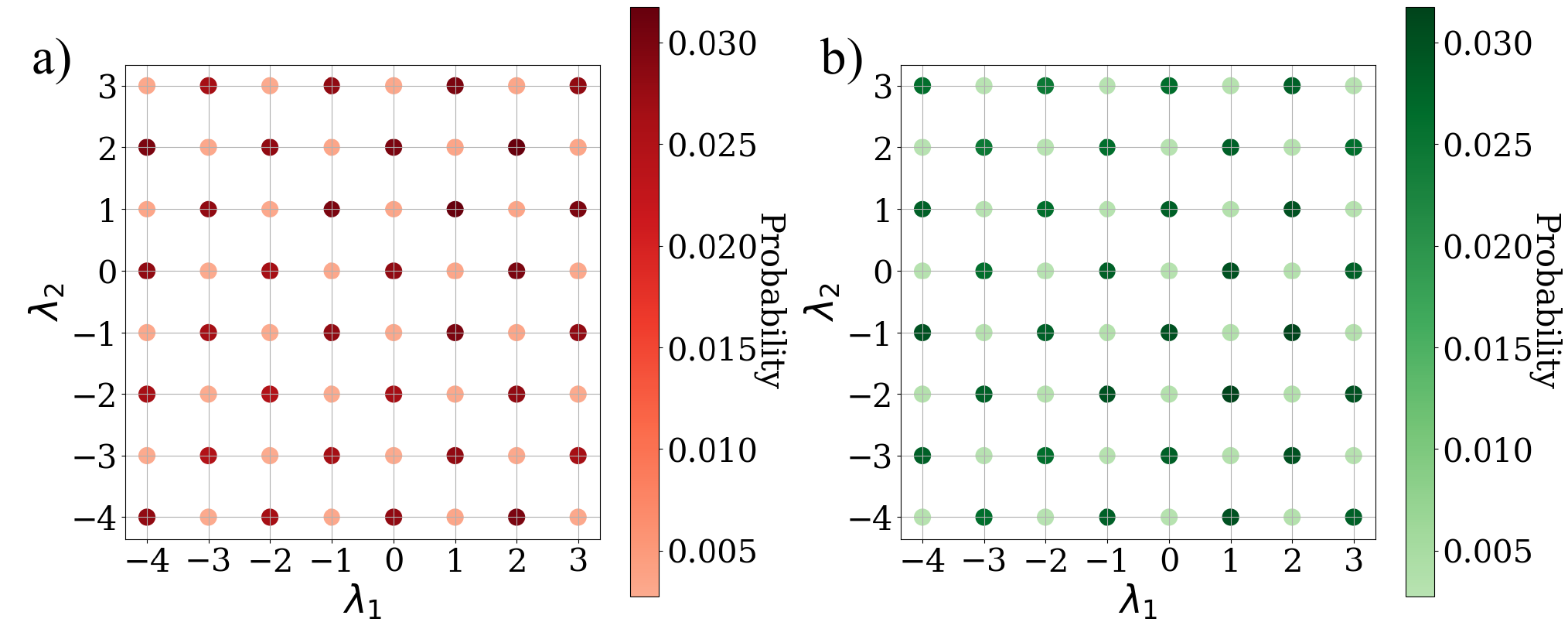}\\[1ex]
\textbf{B)}\\
\FloatBarrier
\includegraphics[width=1\columnwidth]{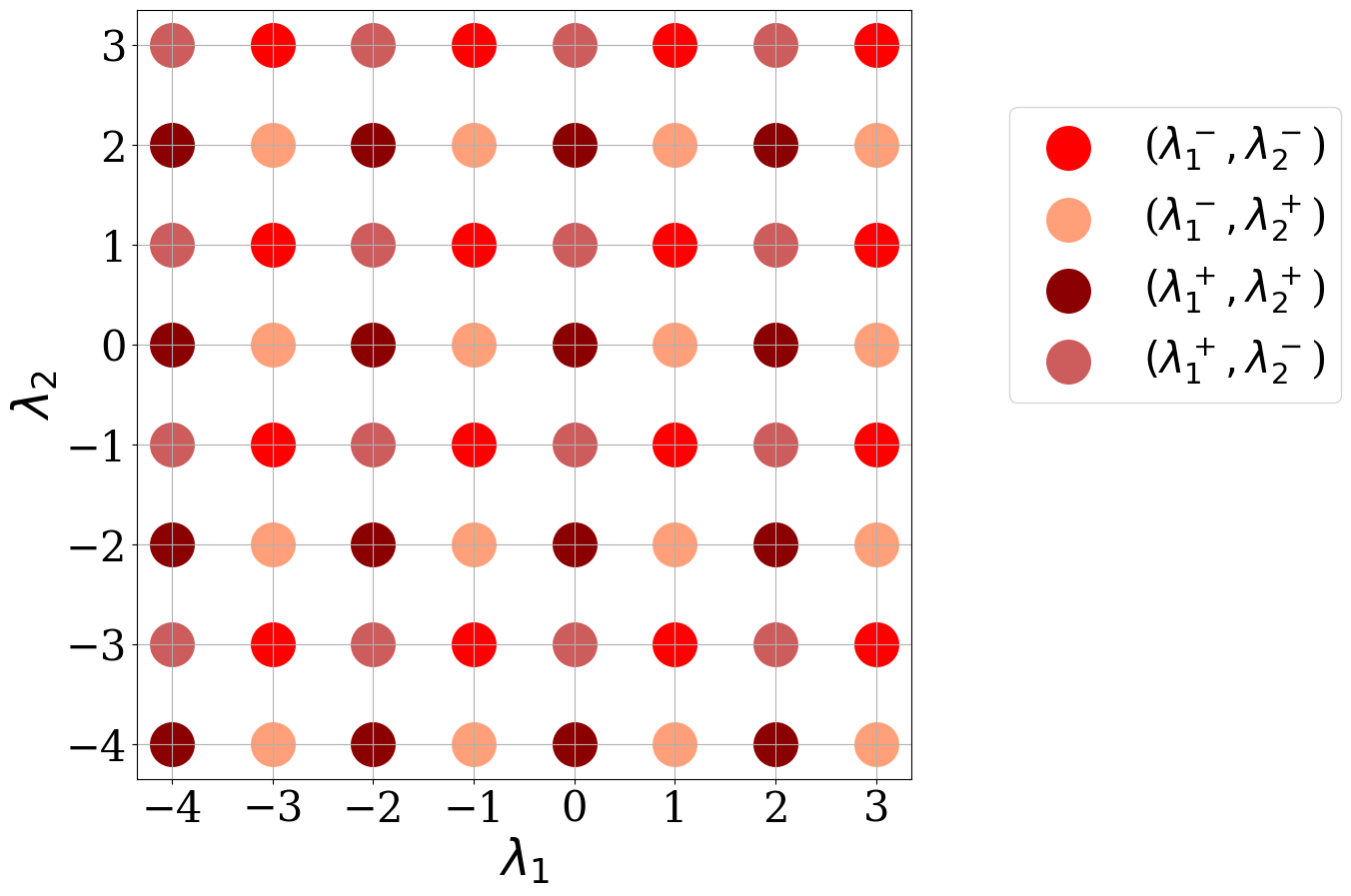}\\[1ex]

\caption{%
\textbf{A)} Probability distributions of the hidden variables $\lambda_{1,2}$
for the different measurement settings: a) $P(\lambda_{1},\lambda_{2}|\hat{a},\hat{b})_{-t}$,
b) $P(\lambda_{1},\lambda_{2}|\hat{a},\hat{b}')_{-t}$. Probabilities
of arrival are $P_{\pm\pm}^{\hat{a},\hat{b}}=P_{\pm\pm}^{\hat{a}',\hat{b}}=P_{\pm\pm}^{\hat{a}',\hat{b}'}=P_{\pm\mp}^{\hat{a},\hat{b}'}=0.45$
and $P_{\pm\mp}^{\hat{a},\hat{b}}=P_{\pm\mp}^{\hat{a}',\hat{b}}=P_{\pm\mp}^{\hat{a}',\hat{b}'}=P_{\pm\pm}^{\hat{a},\hat{b}'}=0.05$,
calculated at $t=10$. 
\textbf{B)} Possible values of $\lambda_{1,2}$ leading
to each of the four different targets $(\lambda_{1}^{\pm},\lambda_{2}^{\pm})$
for the measurement setting pair $(\hat{a},\hat{b})$. \label{fig:prob_dist_sym}}
\end{figure}

Figure \ref{fig:prob_dist_sym} A) depicts the probability distributions
of $\lambda_{1}$ and $\lambda_{2}$ at $t=10$ for a) $P(\lambda_{1},\lambda_{2}|\hat{a},\hat{b})_{-t}$
and b) $P(\lambda_{1},\lambda_{2}|\hat{a},\hat{b}')_{-t}$. A similar
distribution to a) emerges for $P(\lambda_{1},\lambda_{2}|\hat{a}',\hat{b})_{-t}$
and $P(\lambda_{1},\lambda_{2}|\hat{a}',\hat{b'})_{-t}$. It can be
seen that at an even time-step, the higher probabilities in pattern
a) lie at even-even and odd-odd points whilst for b) they lie at even-odd
and odd-even points, a clear difference between $P(\lambda_{1},\lambda_{2}|\hat{a},\hat{b}')_{-t}$
and the other three probability distributions This results in a sufficiently
high additional term to account for the imposed outcome correlations.
Figure \ref{fig:prob_dist_sym} (B) illustrates the possible values
of $\lambda_{1}$ and $\lambda_{2}$ that lead to each target for
settings $\hat{a},\hat{b}$ at $t=10$. A similar plot is formed for
the other settings pairs. Each target is associated with a unique
set of possible prior values of $\lambda_{1}$ and $\lambda_{2}$.
Only even-even values of $\lambda_{1}$ and $\lambda_{2}$ lead to
the even-even targets (at an even time-step) and similarly for other
parities of the targets and $\lambda_{1}$ and $\lambda_{2}$. There
are therefore unique basins of attraction leading to each target for
a given measurement situation.

\begin{figure}[H]
\centering
\includegraphics[width=1\columnwidth]{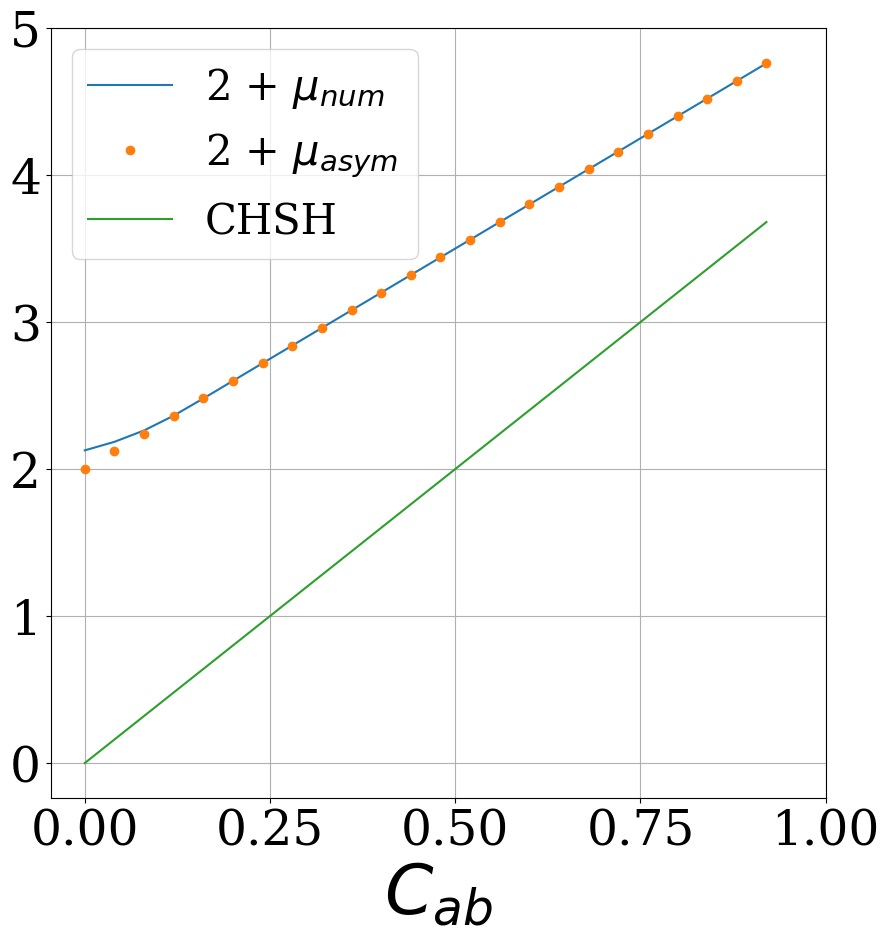}

\caption{The upper bound $2+\mu$ of the CHSH parameter for different values
of the correlation function $C(\hat{a},\hat{b})$ for a symmetric
random walk on the phase space grid with point targets and equal probabilities
of moving one space left or right. $2+\mu_{num}$ is calculated numerically
at $t=10$ (blue curve) and compared with the asymptotic value of
$2+\mu_{asym}$ (orange dots) and the corresponding value of the CHSH
parameter (green). The post-measurement outcome probabilities form
the same pattern as depicted in Figure \ref{fig:Four-targets}, those
with darker shades mean $P_{\pm\pm}^{\hat{a},\hat{b}}=P_{\pm\pm}^{\hat{a}',\hat{b}}=P_{\pm\pm}^{\hat{a}',\hat{b}'}=P_{\pm\mp}^{\hat{a},\hat{b}'}$
take values in the range: $0.25$ to $0.48$ whilst those with the
lighter shades correspond to $P_{\pm\mp}^{\hat{a},\hat{b}}=P_{\pm\mp}^{\hat{a}',\hat{b}}=P_{\pm\mp}^{\hat{a}',\hat{b}'}=P_{\pm\pm}^{\hat{a},\hat{b}'}$
varying from 0.02 to 0.25. \label{fig:sym_walk}}
\end{figure}

Figure \ref{fig:sym_walk} depicts how the numerically calculated
upper bound of the CHSH parameter (Eq. (\ref{eq:1}) (green) depends
on the value of the correlation function $C(\hat{a},\hat{b})$ for
the specified dynamics. The asymptotic value of the upper bound $2+\mu_{asym}$
(orange dots) is plotted alongside the numerically calculated value
$2+\mu_{num}$ (blue dashes), found at time $t=10$. Once again,
$t$ denotes how much time has elapsed between system preparation
and measurement.

It can be seen in Figure \ref{fig:sym_walk} that the additional term
can accommodate the corresponding value of the CHSH parameter. The
asymptotic value of the additional term closely resembles the numerically
calculated value. Naturally, the asymptotic and numerical values
of the additional term become more similar as $t$ increases.

Note that the additional term is nonzero asymptotically only when
the size of the grid is even. When it is odd, the additional term
vanishes as $t\to\infty$ since the the parity of the hidden variable
is not maintained with time and the basins of attraction to each target
mix. We consider another case of the elimination of the additional
term in the next subsection. 

\subsection{Remaining still erases the additional term}

We now consider dynamics with equal probabilities of moving left or
right with step size 1 or of remaining still. The probability of adopting
the initial hidden variable value $\lambda_{i}$ at backward time-step
$t$, given its arrival at a target $\lambda_{i}^{\pm}$ for the measurement
setting $\hat{n}_{i}$ at $t=0$, can be calculating using the following
expression

\begin{equation}
p(\lambda_{i}|\lambda_{i}^{\pm},\hat{n}_{i})_{-t}=\sum_{N_{r},N_{l},N_{s}}\frac{1}{3^{t}}\frac{t!}{N_{r}!N_{l}!N_{s}!}\label{eq:still}
\end{equation}
subject to the condition $N_{r}+N_{l}+N_{s}=t$. $\lambda_{i}^{\pm}$
denotes the target (measurement outcome) for a given measurement setting
$\hat{n}_{i}$, $N_{r}$ denotes the number of moves made to the right,
$N_{l}$ the number of moves made to the left, and $N_{s}$ the number
of times the hidden variable remains the same. The possible initial
values of $\lambda_{i}$ would then be given by $\lambda_{i}=\lambda_{i}^{\pm}+N_{r}-N_{l}$.
The expression reflects that there may be multiple combinations of
values of $N_{r},N_{l}$ and $N_{s}$ that lead to the same value
of $\lambda_{i}$ and thus to yield the total probability of adopting
a specific value of $\lambda_{i}$, we must add the probabilities
of these contributions together. 

\begin{figure}[H]
\centering
\includegraphics[width=1\columnwidth]{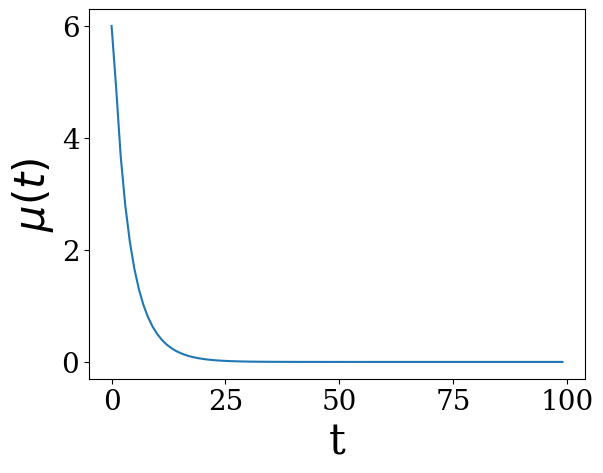}

\caption{The variation of the additional term $\mu$ with $t$ for a symmetric
random walk with equal probabilities of moving left or right or remaining
still. Targets are as depicted in Figure \ref{fig:Four-targets}.
$t$ denotes how much time has elapsed between system preparation
and measurement. The post-measurement outcome probabilities are $P_{\pm\pm}^{\hat{a},\hat{b}}=P_{\pm\pm}^{\hat{a}',\hat{b}}=P_{\pm\pm}^{\hat{a}',\hat{b}'}=P_{\pm\mp}^{\hat{a},\hat{b}'}=0.45$
and $P_{\pm\mp}^{\hat{a},\hat{b}}=P_{\pm\mp}^{\hat{a}',\hat{b}}=P_{\pm\mp}^{\hat{a}',\hat{b}'}=P_{\pm\pm}^{\hat{a},\hat{b}'}=0.05$.
 \label{fig:prob_still} }
\end{figure}

We can use Eq. (\ref{eq:still}) to compute the probability distributions
$p(\lambda_{1}|\lambda_{1}^{\pm},\hat{n}_{1})_{-t}$ and $p(\lambda_{2}|\lambda_{2}^{\pm},\hat{n}_{2})_{-t}$
and Eq. (\ref{eq:calculate pdfs}) provides the pdf $P(\lambda_{1},\lambda_{2}|\hat{n}_{1},\hat{n}_{2})_{-t}$.
We can then calculate how the additional term $\mu$ varies with time
using Eq. (\ref{eq:7}), shown in Figure \ref{fig:prob_still}for
a specific set of post-measurement outcome probabilities $P_{\pm\pm}^{\hat{a},\hat{b}}$
etc. It can be seen that it converges towards zero. Therefore these
dynamics could not have been used to account for the asserted variation
in correlation functions and CHSH parameter, the reason for which
is the failure to maintain separate basins of attraction.

\begin{figure}[H]
\centering
\textbf{A)}\\

\includegraphics[width=1\columnwidth]{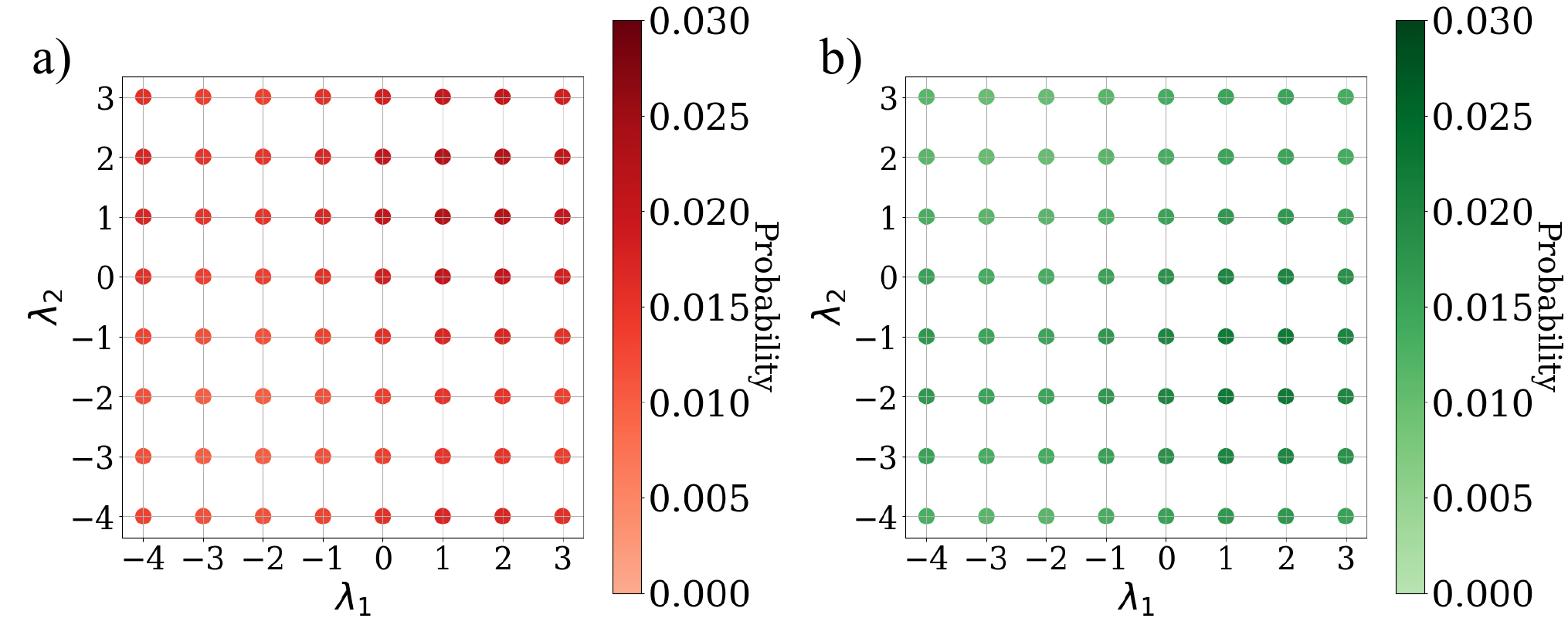}\\[1ex]

\textbf{B)}\\
\FloatBarrier

\includegraphics[width=1\columnwidth]{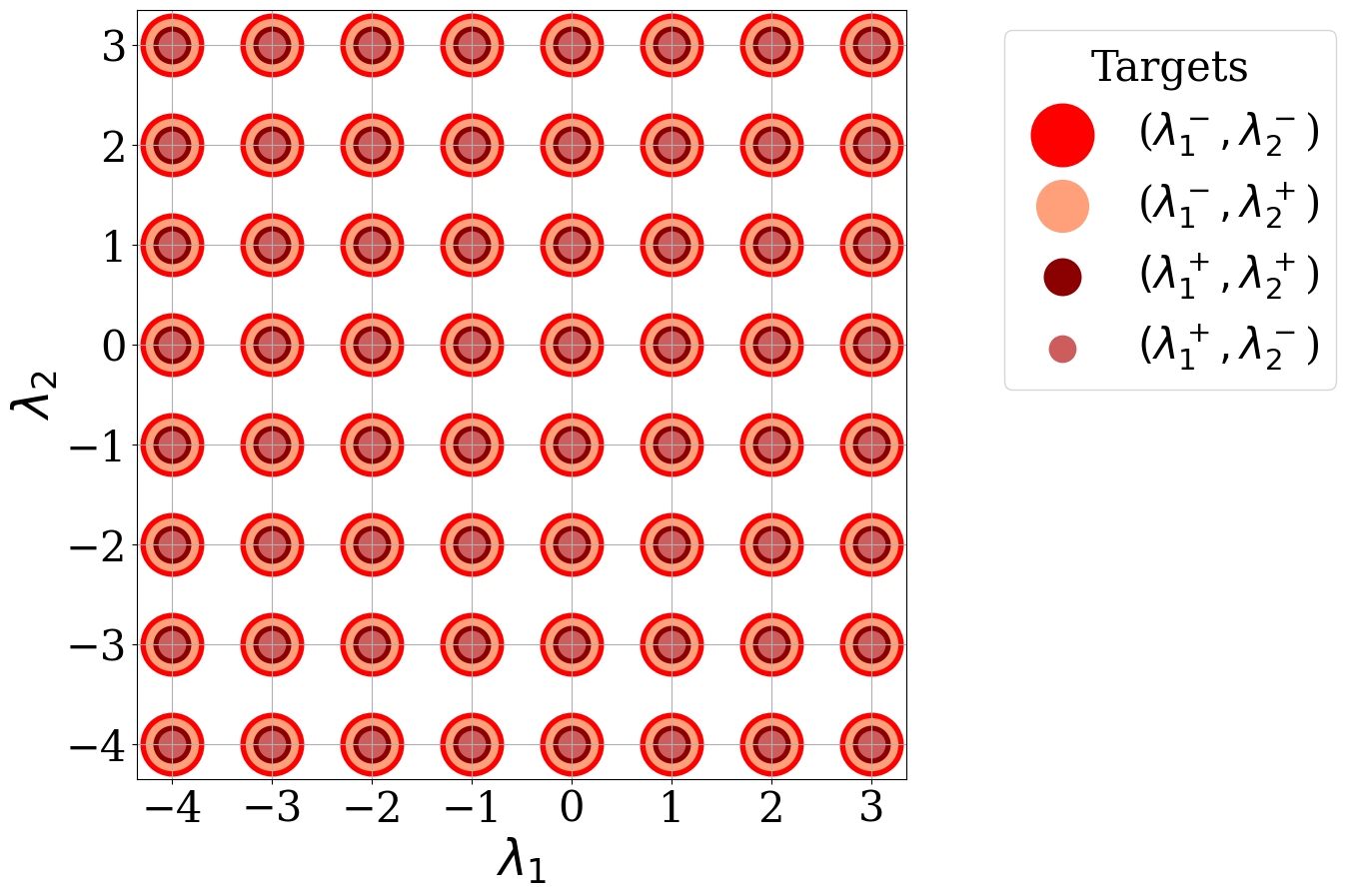}\\[1ex]

\caption{%
\textbf{A)} Probability distributions of the hidden variables at $t=10$ 
a) $P(\lambda_{1},\lambda_{2}|\hat{a},\hat{b})_{-t}$ and b) $P(\lambda_{1},\lambda_{2}|\hat{a},\hat{b}')_{-t}$
for a random walk with equal probabilities of remaining still or moving
left or right by one spacing.  The post-measurement outcome probabilities
are $P_{\pm\pm}^{\hat{a},\hat{b}}=P_{\pm\pm}^{\hat{a}',\hat{b}}=P_{\pm\pm}^{\hat{a}',\hat{b}'}=P_{\pm\mp}^{\hat{a},\hat{b}'}=0.45$
and $P_{\pm\mp}^{\hat{a},\hat{b}}=P_{\pm\mp}^{\hat{a}',\hat{b}}=P_{\pm\mp}^{\hat{a}',\hat{b}'}=P_{\pm\pm}^{\hat{a},\hat{b}'}=0.05$.
\textbf{B)} The possible values of $\lambda_{1}$ and $\lambda_{2}$ that
lead to each of the four different targets $(\lambda_{1}^{\pm},\lambda_{2}^{\pm})$
for setting $(\hat{a},\hat{b})$. \label{fig:prob_dist_prob_still} }
\end{figure}

Figure \ref{fig:prob_dist_prob_still} (A) depicts the probability
distributions of the hidden variables at $t=10$ for (a) $P(\lambda_{1},\lambda_{2}|\hat{a},\hat{b})_{-t}$
and (b) $P(\lambda_{1},\lambda_{2}|\hat{a},\hat{b}')_{-t}$. Unlike
the symmetric random walk with equal probabilities of moving left
and right, adding an option of remaining generates probability distributions
with similar patterns for the different settings. 

Figure \ref{fig:prob_dist_prob_still} (B) shows possible values of
$\lambda_{1}$ and $\lambda_{2}$ associated with each target for
the measurement setting pair $(\hat{a},\hat{b})$. Similar plots are
formed for the other three measurement settings. All targets appear
to be accessible from all possible values of $\lambda_{1}$ and $\lambda_{2}$.
There are therefore no unique basins of attraction and therefore very
little difference between the hidden variable distributions in Figure
\ref{fig:prob_dist_prob_still} (A). Consequently there is no measurement
dependence the additional term in Figure \ref{fig:prob_still} goes
asymptotically to zero. 

\subsection{Broad targets}

\begin{figure}[H]
\centering
\includegraphics[width=1\columnwidth]{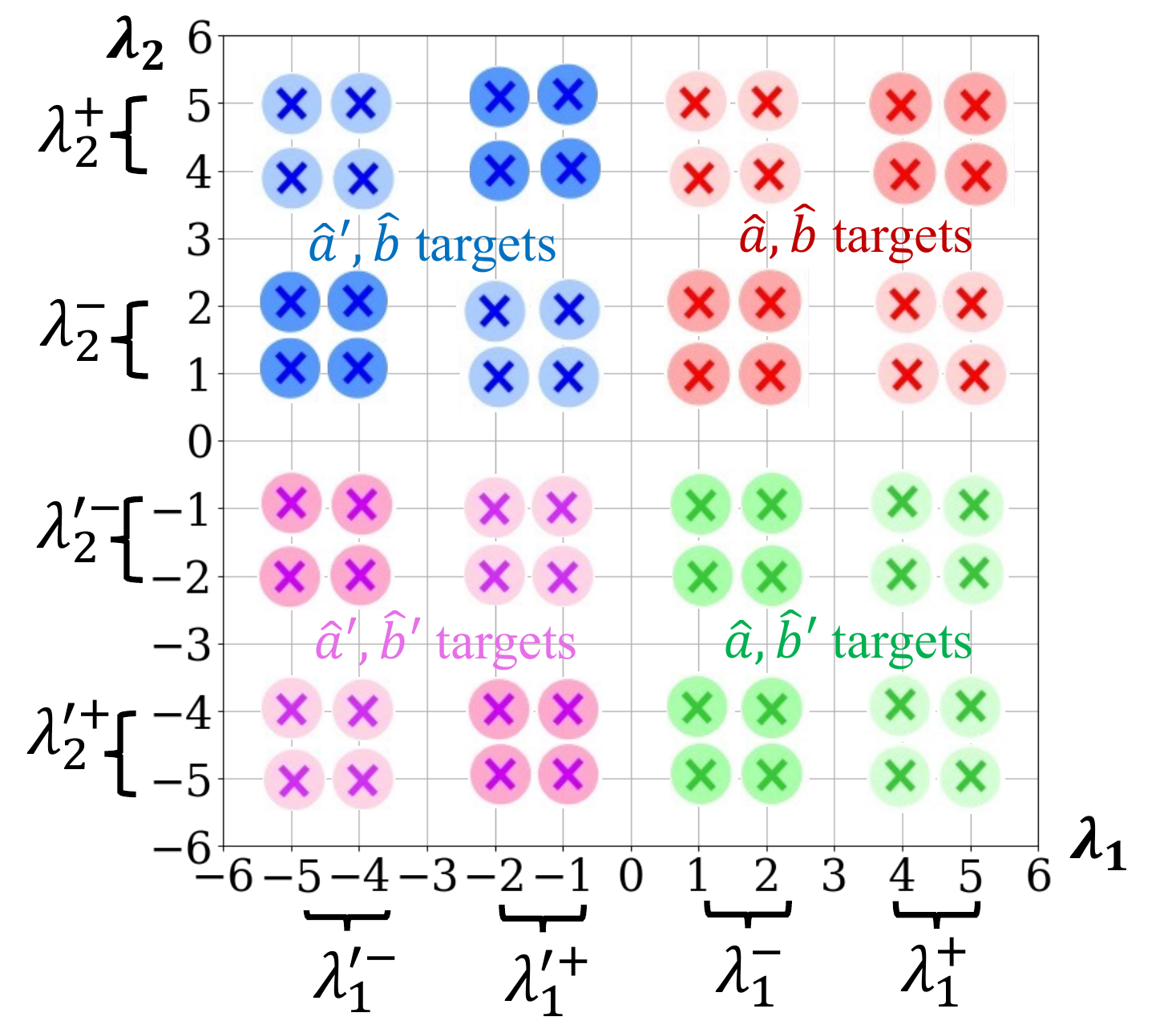}

\caption{Broad targets (measurement outcomes) comprised of four points are
used for each measurement setting. The darker the shade, the higher
the post-measurement outcome probability. As before, $P_{\pm\pm}^{\hat{a},\hat{b}}=P_{\pm\pm}^{\hat{a}',\hat{b}}=P_{\pm\pm}^{\hat{a}',\hat{b}'}=P_{\pm\mp}^{\hat{a},\hat{b}'}=0.45$
and $P_{\pm\mp}^{\hat{a},\hat{b}}=P_{\pm\mp}^{\hat{a}',\hat{b}}=P_{\pm\mp}^{\hat{a}',\hat{b}'}=P_{\pm\pm}^{\hat{a},\hat{b}'}=0.05$,
with probability distributed evenly across the four points forming
each target. \label{fig:broad_targets}}
\end{figure}

So far we have considered the targets (arrival at which signifies
measurement outcomes) to be single points in the phase space of the
hidden variables, characterised by their parity. Now we consider them
to correspond to a collection of points that includes values of $\lambda_{1}$
and $\lambda_{2}$ of all parities. Figure \ref{fig:broad_targets}
illustrates how each target is now broadened to comprise four points:
an even-even, odd-odd, even-odd and an odd-even point. The post-measurement
outcome probabilities for each target are divided equally between
the four points. Measurement setting $(\hat{a},\hat{b})$ now has
a $A=-1$, $B=-1$ target $(\lambda_{1}^{-},\lambda_{2}^{-})$ comprised
of the following four points $(\lambda_{1},\lambda_{2})=(1,1),(1,2),(2,2),(2,1)$,
a $(\lambda_{1}^{+},\lambda_{2}^{+})$ target comprised of $(4,4),(4,5),(5,5),(5,4)$,
a $(\lambda_{1}^{+},\lambda_{2}^{-})$ target made up of $(4,1),(4,2),(5,2),(5,1)$
and a $(\lambda_{1}^{-},\lambda_{2}^{+})$ target that includes $(1,4),(1,5),(2,4),(2,5)$,
with a similar pattern for the other three measurement axis pairs
in their respective quadrants. The initial probability distributions
over the values of $\lambda_{1}$ and $\lambda_{2}$ will be conditioned
on later arrival at any of the four points within each target. We
calculate the probability distributions $P(\lambda_{1},\lambda_{2}|\hat{n}_{1},\hat{n}_{2})_{-t}$
and hence the additional term $\mu$ using Eq. (\ref{eq:7}). We use
the periodic boundaries of size 12 defined by $\lambda=((\lambda+6)\oplus12)-6$.

\begin{figure}[H]
\centering
\includegraphics[width=1\columnwidth]{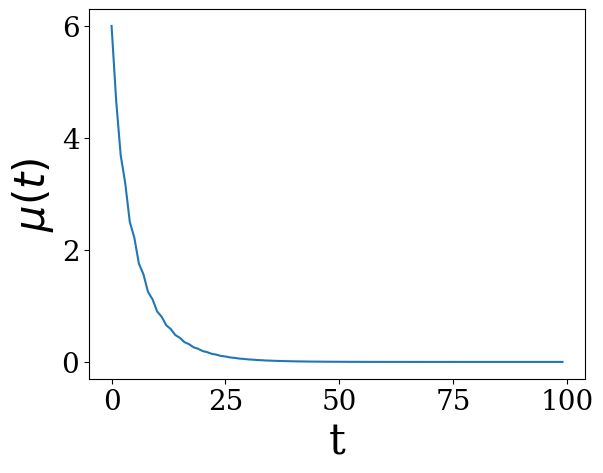}

\caption{The variation of the additional term $\mu$ with $t$ for a symmetric
random walk with equal probability of  moving left or right by one
spacing and broad targets which are as depicted in Figure \ref{fig:broad_targets}.
The outcome probabilities are $P_{\pm\pm}^{\hat{a},\hat{b}}=P_{\pm\pm}^{\hat{a}',\hat{b}}=P_{\pm\pm}^{\hat{a}',\hat{b}'}=P_{\pm\mp}^{\hat{a},\hat{b}'}=0.45$
and $P_{\pm\mp}^{\hat{a},\hat{b}}=P_{\pm\mp}^{\hat{a}',\hat{b}}=P_{\pm\mp}^{\hat{a}',\hat{b}'}=P_{\pm\pm}^{\hat{a},\hat{b}'}=0.05$.
 \label{fig:broad_targs_Res}}
\end{figure}

Figure \ref{fig:broad_targs_Res} illustrates how the additional term
evolves over time $t$ running backwards from system measurement to
preparation for a symmetric random walk with equal probabilities of
moving left or right by one spacing and broad targets. The additional
term decreases asymptotically towards zero indicating that such dynamics
and targets are incompatible with assumed outcome probabilities that
break the standard Bell bound. 

\begin{figure}[H]
\centering
\includegraphics[width=1\columnwidth]{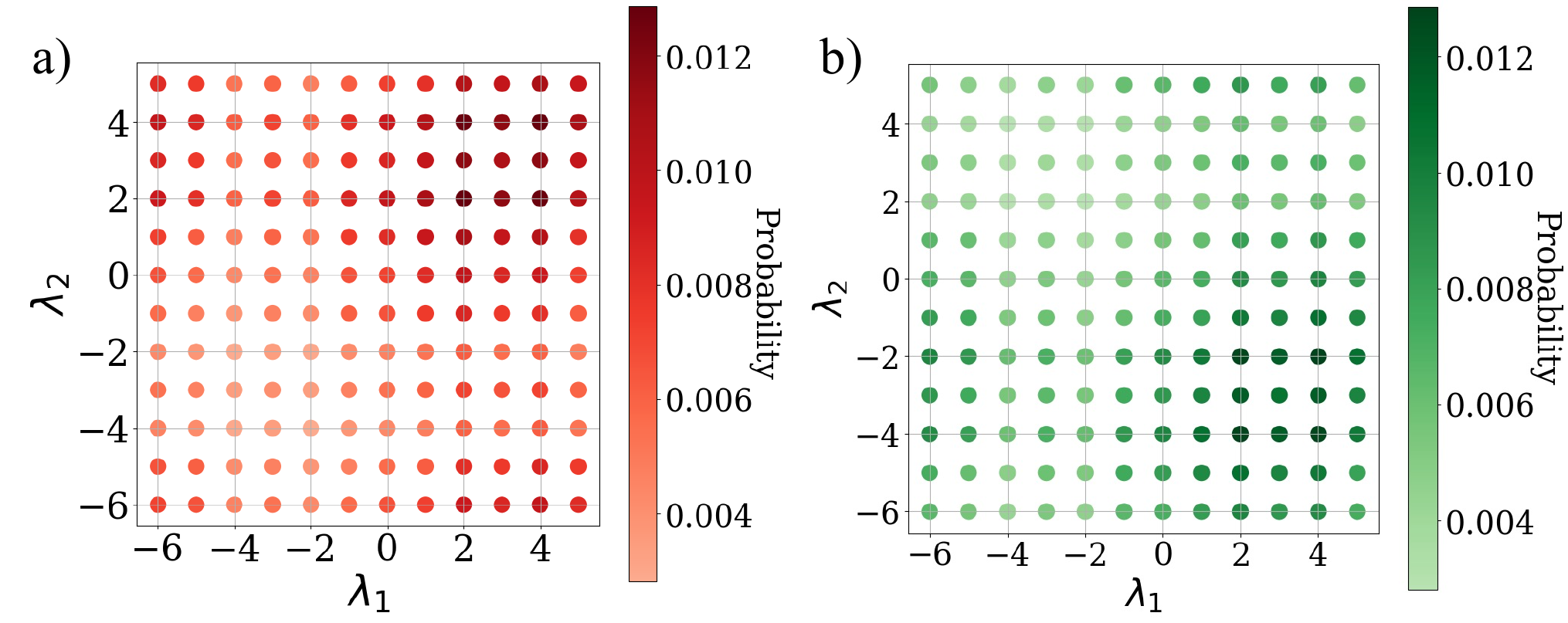}

\caption{The probability distributions of the hidden variables $\lambda_{1}$
and $\lambda_{2}$ at $t=10$: (a) $P(\lambda_{1},\lambda_{2}|\hat{a},\hat{b})_{-t}$
and (b) $P(\lambda_{1},\lambda_{2}|\hat{a},\hat{b}')_{-t}$ for a
symmetric random walk with equal probability of moving left or right
by one spacing and broad targets. The outcome probabilities are $P_{\pm\pm}^{\hat{a},\hat{b}}=P_{\pm\pm}^{\hat{a}',\hat{b}}=P_{\pm\pm}^{\hat{a}',\hat{b}'}=P_{\pm\mp}^{\hat{a},\hat{b}'}=0.45$
and $P_{\pm\mp}^{\hat{a},\hat{b}}=P_{\pm\mp}^{\hat{a}',\hat{b}}=P_{\pm\mp}^{\hat{a}',\hat{b}'}=P_{\pm\pm}^{\hat{a},\hat{b}'}=0.05$.
 \label{fig:prob_dist_broad_targ}}
\end{figure}

Figure \ref{fig:prob_dist_broad_targ} depicts the probability distributions
over $\lambda_{1}$ and $\lambda_{2}$ for (a) measurement setting
$(\hat{a},\hat{b})$ and (b) $(\hat{a},\hat{b}')$ at $t=10$. As
in Figure \ref{fig:prob_dist_prob_still} (A), there are no significant
differences between the probability distributions for the different
measurement settings. There are no unique basins of attraction to
each target and the additional term vanishes asymptotically. Measurement
dependence is absent.

\section{Conclusions \label{sec:Conclusions}}

Bell's inequalities rely on an assumption that the probability density
$\rho$ of hidden variables $\lambda$ prior to measurement does not
depend on the measurement settings chosen, a condition known as measurement
independence. If in a Bell experiment particle 1 undergoes a measurement
of its spin along either axis $\hat{a}$ or $\hat{a}'$ and particle
2 undergoes a spin measurement along axis $\hat{b}$ or $\hat{b}'$,
then measurement independence is expressed by $\rho(\lambda|\hat{a},\hat{b})=\rho(\lambda|\hat{a}',\hat{b})=\rho(\lambda|\hat{a},\hat{b}')=\rho(\lambda|\hat{a}',\hat{b}')$.
In this work we relax this assumption by imagining that measurement
is a dynamic process involving attraction of the system towards fixed
points in the hidden variable phase space that represent measurement
outcomes. A suitable measurement dynamics together with fixed points
that are specific to measurement settings can then lead to measurement
dependence. We demonstrate that the upper bound of the CHSH parameter
may be raised by an additional term $\mu$, defined by the sums and
differences between the four hidden variable probability densities
conditioned on the measurement settings. This can be large enough
to accommodate measurement correlations that break the standard Bell
bound. 

The conditioned probability distributions of the hidden variables
$\rho(\lambda|\hat{n}_{1},\hat{n}_{2})$ need not correspond to a
distribution from which the hidden variables might be selected to
represent a prepared but uncertain entangled state. Rather, they demonstrate
what the hidden variable statistics before measurement would have
had to have been in order to account for the Bell bound breaking measurement
correlations. Our reasoning works backward from measurement outcomes
rather than forward from the selection of initial hidden variables.
If the outcomes of measurements are encoded in the hidden variables,
we may infer the probability density of the hidden variables before
measurement, given a set of measurement settings, an associated distribution
of measurement outcomes, and a model of the dynamics associated with
measurement. 

We consider a toy model where the dynamics of the hidden variables
take the form of a symmetric random walk on a discrete space with
periodic boundaries, and with equal probabilities of moving left or
right. Variations in the dynamics are also explored. Measurement drives
the hidden variables towards post-measurement attractors (targets)
defining measurement outcomes for the given measurement settings.
We are able to infer the probability distributions of the hidden variables
before measurement, given the chosen measurement settings and explore
how the additional term $\mu$ evolves with the amount of time between
state preparation and measurement. Variations in the dynamics are
explored to reveal key features required for measurement dependence
to be significant. 

The additional term depends on three features: the dynamics of the
random walk, the size of the periodic boundaries and the level of
coarse graining of the measurement outcomes (targets). The dynamics
that successfully reproduce Bell bound breaking correlations are therefore
delicate. What is required is that each target for a given measurement
situation has to some extent a basin of attraction under the dynamics.
We show this by introducing an additional probability of remaining
still (as well as moving left or right) into the random walk dynamics,
as well as coarse graining the targets such that they comprised multiple
points on the phase space, which disturb the basins of attraction
offered by the simple random walk.

Not unlike contextuality, we learn that in order to break the Bell
bound it is necessary that only a subset of the possible values of
the hidden variables should be capable of providing certain measurement
outcomes for a specified measurement scenario. \medskip{}

\section*{Acknowledgements}

SMW is supported by a PhD studentship funded by Engineering and Physical
Sciences Research Council (EPSRC) under grant codes EP/R513143/1 and
EP/T517793/1.\bibliography{combined_mes_dep_paper_5_IJF}

\appendix

\section{Symmetric random walk with an unequal probability of remaining still
or moving left or right with step size 2\label{sec:Symmetric-random-walk}}

We now consider the case of a symmetric random walk with a probability
of $\frac{1}{4}$ of moving left or right with step size 2: $\Delta\lambda_{1}=\Delta\lambda_{2}=2$
and a probability $\frac{1}{2}$ of remaining still.  The probability
of adopting a particular value of $\lambda_{i}$ at time $-t$, given
the system has collapsed at measurement outcome (target) $\lambda_{i}^{\pm}$
associated with the measurement setting $\hat{n}_{i}$ at $t=0$,
is then expressed as:

\begin{widetext}

\begin{equation}
P(\lambda_{i}|\lambda_{i}^{\pm},\hat{n}_{i})_{-t}=\sum_{N_{r},N_{l},N_{s}}\left(\frac{1}{4}\right)^{N_{r}}\left(\frac{1}{4}\right)^{N_{l}}\left(\frac{1}{2}\right)^{N_{s}}\frac{t!}{N_{r}!N_{l}!N_{s}!},\label{eq:still-1}
\end{equation}

\end{widetext}subject to the restriction $t=N_{s}+N_{r}+N_{l}$,
where $N_{r}$ is the total number of times the system moves right
and $N_{l}$ the total number of times it moves left in a given time
$t$. $N_{s},N_{r}$ and $N_{l}$ can all take integer values between
$0\rightarrow t$. The possible values of $\lambda_{i}$ would then
be given by $\lambda_{i}=\lambda_{i}^{\pm}+N_{r}-N_{l}$. The expression
in the sum in Eq. (\ref{eq:still-1}) reflects that there may be multiple
combinations of values of $N_{r},N_{l}$ and $N_{s}$ that lead to
the same value of $\lambda_{i}$ and thus to compute the total probability
of adopting the value $\lambda_{i}$, we must add these contributions
together. The probability distribution over $\lambda_{1}$ and $\lambda_{2}$
can then be found through Eq. (\ref{eq:calculate pdfs}), since the
random walks in each dimension are independent. We can calculate the
additional term $\mu$ in the Bell bound using Eq. (\ref{eq:7}). 

\begin{figure}[H]
\centering
\includegraphics[width=1\columnwidth]{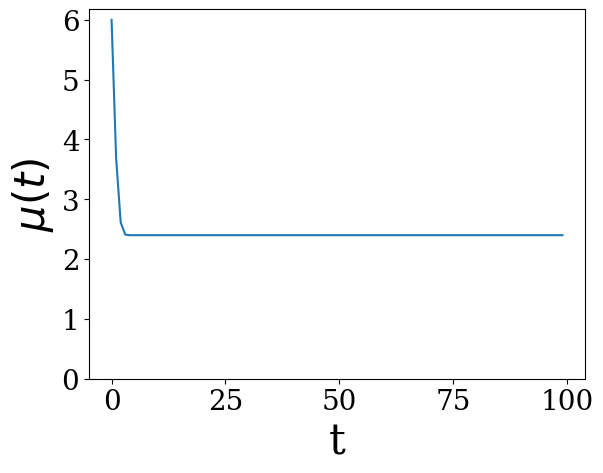}

\caption{The evolution of the additional term $\mu(t)$ where $t$ is the time
between when the system is measured and when it was prepared. The
dynamics are such that there is a probability of $\frac{1}{4}$ of
moving left or right with step size 2: $\Delta\lambda_{1}=\Delta\lambda_{2}=2$
and a probability $\frac{1}{2}$ of remaining still. The probabilities
of arrival are $P_{\pm\pm}^{\hat{a},\hat{b}}=P_{\pm\pm}^{\hat{a}',\hat{b}}=P_{\pm\pm}^{\hat{a}',\hat{b}'}=P_{\pm\mp}^{\hat{a},\hat{b}'}=0.45$
and $P_{\pm\mp}^{\hat{a},\hat{b}}=P_{\pm\mp}^{\hat{a}',\hat{b}}=P_{\pm\mp}^{\hat{a}',\hat{b}'}=P_{\pm\pm}^{\hat{a},\hat{b}'}=0.05$.
\label{fig:prob_2_p_still_mu_over_time}}
\end{figure}

Figure \ref{fig:prob_2_p_still_mu_over_time} illustrates how the
numerically calculated additional term $\mu$ varies with time $t$
for the aforementioned dynamics. It can be seen that the additional
term decreases to a plateau at a value of around 2.4, where it remains,
similarly to the dynamics of a symmetric random walk with equal probability
of moving left or right by one spacing. At $t=100$, the additional
term remains nonzero, revealing that the conditioning of the probability
distributions of the hidden variables on the chosen measurement setting
persists.

\begin{figure}[H]
\centering
\includegraphics[width=1\columnwidth]{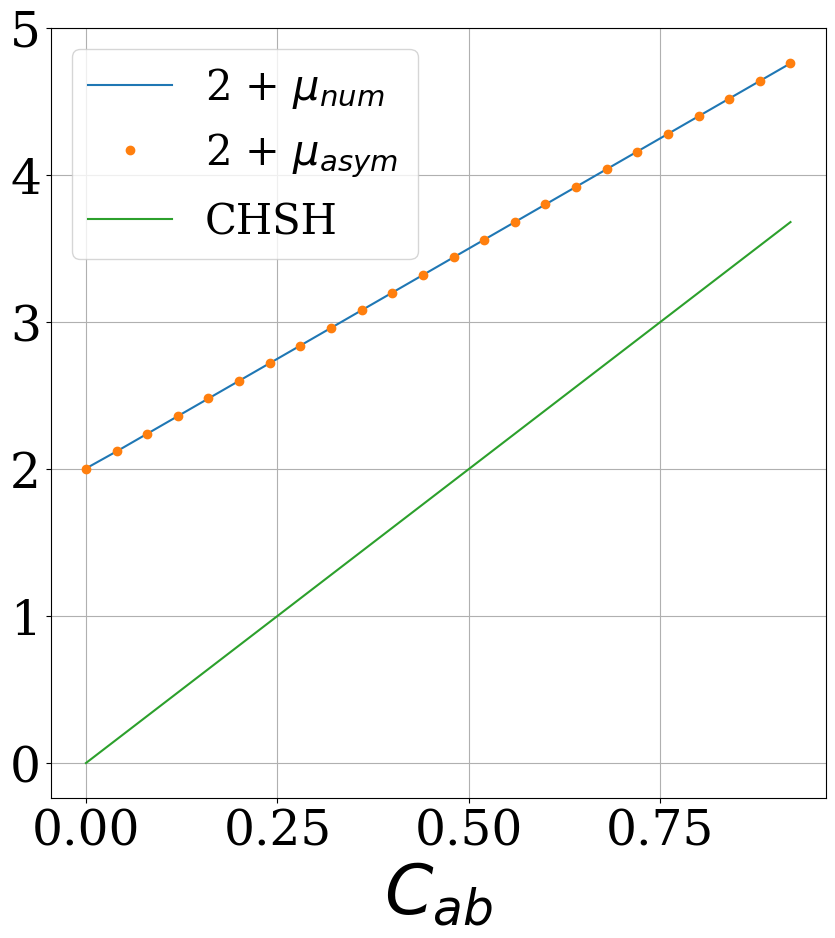}

\caption{The new upper bound $2+\mu$ of the CHSH parameter for different values
of the correlation function $C(\hat{a},\hat{b})$ for a symmetric
random walk with point targets and a probability of $\frac{1}{4}$
of moving left or right with step size 2 and a probability of $\frac{1}{2}$
of remaining still. $2+\mu_{num}$ is calculated numerically at $t=10$
(blue dashes) and compared with the asymptotic value $2+\mu_{asym}$
(orange dots) and the corresponding Bell bound on the CHSH parameter
(green). The probabilities of arrival form the same pattern as depicted
in Figure \ref{fig:Four-targets}. \label{fig:2_p_still_add_term_Corr}}
\end{figure}

Figure \ref{fig:2_p_still_add_term_Corr} depicts how the numerical
$2+\mu_{num}$ (blue dashes) and asymptotically calculated $2+\mu_{asym}$
(orange dots) additional term vary with $C(\hat{a},\hat{b})$, and
contrasted with the Bell bound on the CHSH parameter (green).  The
asymptotic value of the additional term is found in a similar way
to that outlined in Section \ref{subsec:analytics}. Here it appears
to exactly match the numerically calculated value. 

\begin{figure}[H]
\centering
\textbf{A)}\\

\includegraphics[width=1\columnwidth]{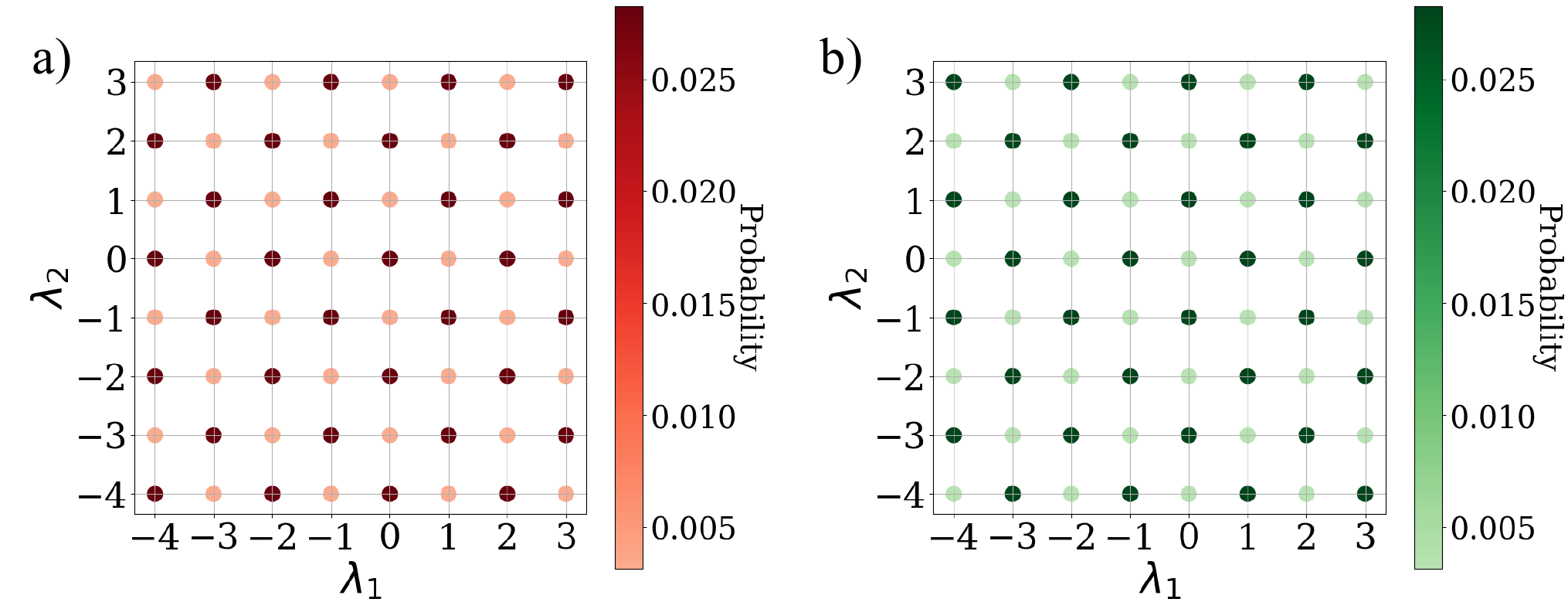}\\[1ex]

\textbf{B)}\\
\FloatBarrier

\includegraphics[width=1\columnwidth]{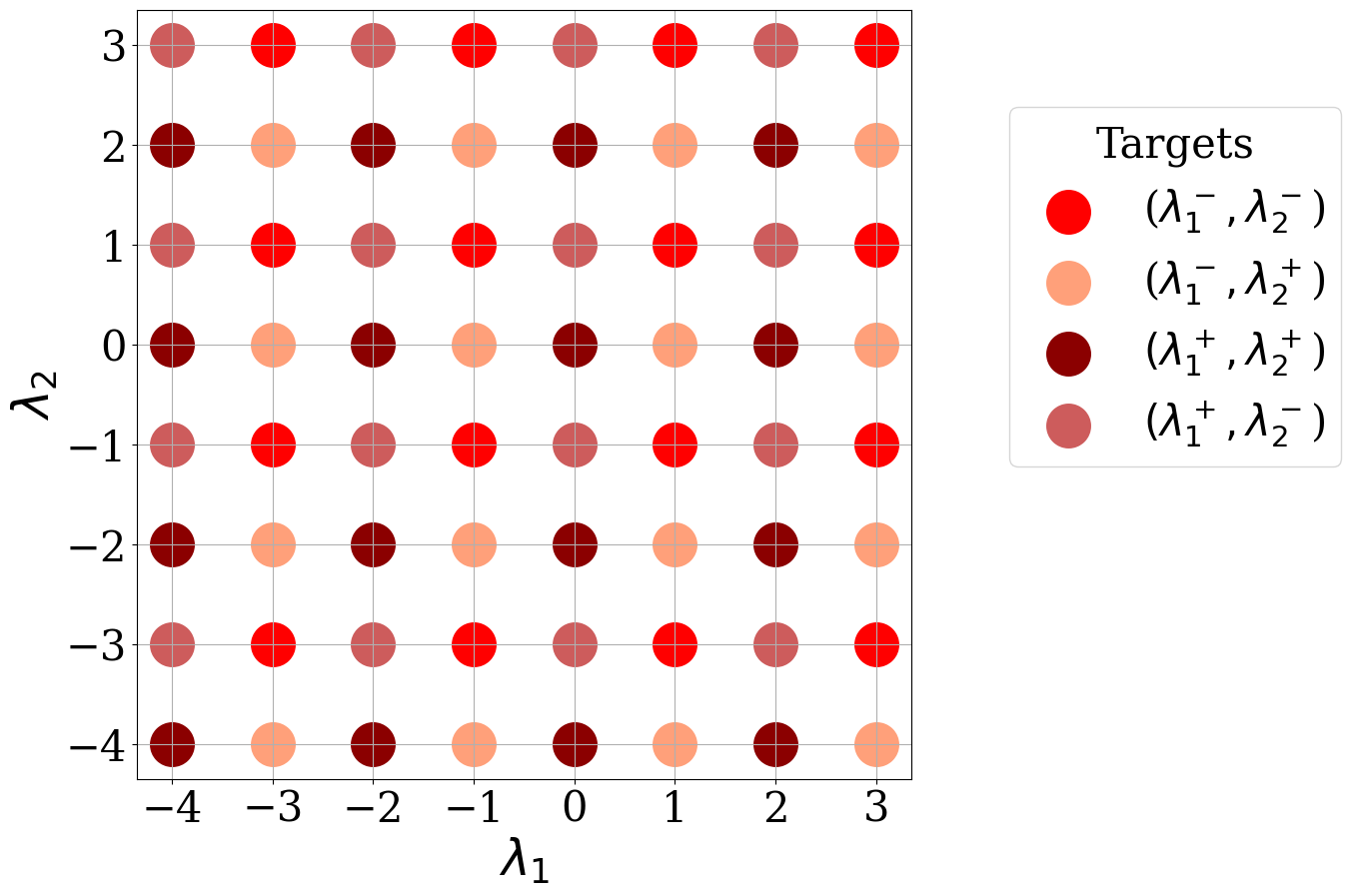}\\[1ex]

\caption{%
\textbf{A)} The evolution of the probability distribution of the hidden variables
for measurement setting (a) $(\hat{a},\hat{b})$ and b) $(\hat{a},\hat{b}')$
at time $t=10$ and with probabilities of arrival $P_{\pm\pm}^{\hat{a},\hat{b}}=P_{\pm\pm}^{\hat{a}',\hat{b}}=P_{\pm\pm}^{\hat{a}',\hat{b}'}=P_{\pm\mp}^{\hat{a},\hat{b}'}=0.45$
and $P_{\pm\mp}^{\hat{a},\hat{b}}=P_{\pm\mp}^{\hat{a}',\hat{b}}=P_{\pm\mp}^{\hat{a}',\hat{b}'}=P_{\pm\pm}^{\hat{a},\hat{b}'}=0.05$.
\textbf{B)} The possible values of $\lambda_{1}$ and $\lambda_{2}$ associated
with arrival at each of the four targets when $\hat{a}$ and $\hat{b}$
are chosen to be the measurement settings at $t=10$. \label{fig:2_p_still_prob_paths}}
\end{figure}

Figure \ref{fig:2_p_still_prob_paths}A) depicts how the probability
distribution of the hidden variables varies for measurement settings
(a) $(\hat{a},\hat{b})$ and b) $(\hat{a},\hat{b}')$. It can be seen
that the probability for odd-odd and even-even values of $\lambda_{1}$,$\lambda_{2}$
is high in Figure \ref{fig:2_p_still_prob_paths}A a) while in Figure
\ref{fig:2_p_still_prob_paths}A b) it is low. This difference between
the distributions leads to an additional term that is high enough
to account for Bell bound breaking correlations. 

Figure \ref{fig:2_p_still_prob_paths}B) shows the possible values
of $\lambda_{1}$ and $\lambda_{2}$ associated with arrival at each
of the four targets for measurement setting $(\hat{a},\hat{b})$.
For example, only odd-odd values of $\lambda_{1}$,$\lambda_{2}$
are associated with odd-odd targets, and similarly for the other targets.
Only one of the four possible targets may be reached from a particular
value of $\lambda_{1}$ and $\lambda_{2}$, defined by their parity,
such that basins of attraction to each target exist. This is why the
dynamics in question are able to produce sufficiently different probability
distributions for $\lambda_{1}$ and $\lambda_{2}$, such as those
depicted in Figure \ref{fig:2_p_still_prob_paths}A), that can lead
to a nozero asymptotic additional term.

Note that this case is similar to that of the symmetric random walk
dynamics with equal probabilities of moving left or right by one spacing,
but the dynamics are such that at even and odd time-steps the possible
values of $\lambda_{1}$ and $\lambda_{2}$ associated with an even-even
target will both be even and those associated with an odd-even target
will be odd and even respectively, where for the simpler dynamics
this would only be the case when the time-step is even.

\section{Asymmetric random walk dynamics \label{sec:Asymmetric-random-walk}}

In the main text as well as the preceding Appendix, we assumed symmetric
random walk measurement dynamics for both hidden variables. We extend
this to explore a random walk with an asymmetric step size for both
variables: the step size when moving left is $\Delta\lambda_{l}=2$
whilst the step size when moving right is $\Delta\lambda_{r}=1$.
The possible values of $\lambda_{i}$ would then be $\lambda_{i}=\lambda_{i}^{\pm}+\Delta\lambda_{r}N_{r}-\Delta\lambda_{l}(t-N_{r})$
where $N_{r}$ is the total number of moves taken to the right for
a given evolution pathway, taking integer values between 0 and $t$,
and $N_{l}=t-N_{r}$ is the total number of moves taken to the left.
$\lambda_{i}^{\pm}$ denotes the target (measurement outcome) for
a given measurement setting $\hat{n}_{i}$. The probability of adopting
the initial hidden variable value $\lambda_{i}$ is then found using 

\begin{equation}
p(\lambda_{i}|\lambda_{i}^{\pm},\hat{n}_{i})_{-t}=\frac{1}{2^{t}}\frac{t!}{(\lambda_{i}^{\pm}-1)!(t+1-\lambda_{i}^{\pm})!},\label{eq:sym}
\end{equation}
and Eq. (\ref{eq:calculate pdfs}) can then be used to calculate the
probability distributins. We implement periodic boundaries expressed
by $\lambda_{PB}=((\lambda+\frac{9}{2})\oplus9)-\frac{9}{2}$ rather
than $\lambda_{PB}=((\lambda+4)\oplus8)-4$ used previously. The size
of the boundary needs to be a multiple of three in order for the basins
of attraction leading to the four targets in a given measurement situation
not to mix.  

\begin{figure}[H]
\centering
\includegraphics[width=1\columnwidth]{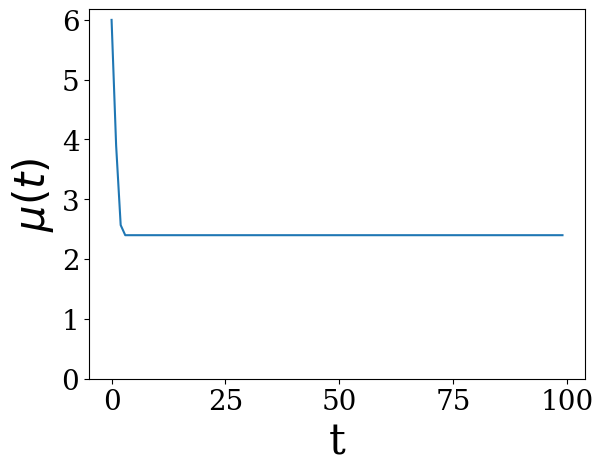}

\caption{The evolution of the numerically calculated additional term $\mu(t)$
where $t$ is the time between when the system was measured and when
it was prepared. The dynamics are such that there is an equal probability
of moving left with step size 2 or right with step size l. The probabilities
of arrival are $P_{\pm\pm}^{\hat{a},\hat{b}}=P_{\pm\pm}^{\hat{a}',\hat{b}}=P_{\pm\pm}^{\hat{a}',\hat{b}'}=P_{\pm\mp}^{\hat{a},\hat{b}'}=0.45$
and $P_{\pm\mp}^{\hat{a},\hat{b}}=P_{\pm\mp}^{\hat{a}',\hat{b}}=P_{\pm\mp}^{\hat{a}',\hat{b}'}=P_{\pm\pm}^{\hat{a},\hat{b}'}=0.05$.\label{fig:asymmetric_mu_with_t}}

\end{figure}

Figure \ref{fig:asymmetric_mu_with_t} illustrates how the additional
term $\mu(t)$ varies with time for the dynamics described above.
Similarly to Figure \ref{fig:prob_2_p_still_mu_over_time}, the additional
terms decreases until it reaches a plateau at a value of around 2.4,
where it remains. It therefore takes a significant value even at asymptotic
time.

\begin{figure}[H]
\centering
\includegraphics[width=1\columnwidth]{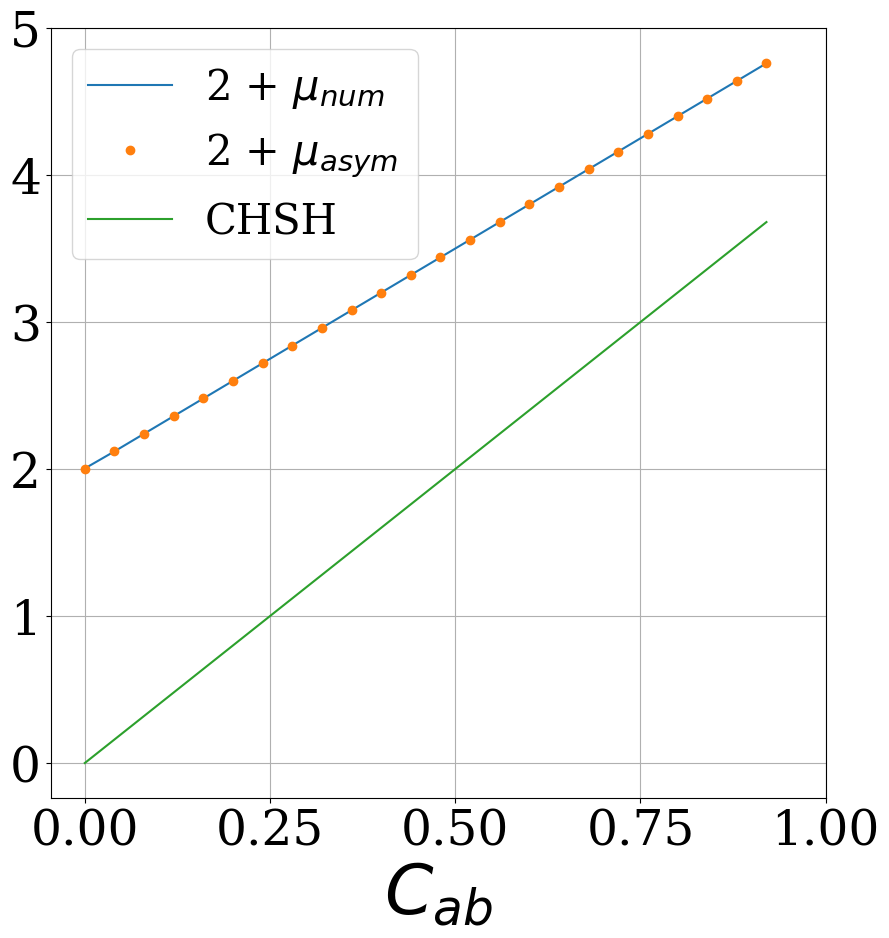}

\caption{The new upper bound $2+\mu$ of the CHSH parameter for different values
of the correlation function $C(\hat{a},\hat{b})$ for a asymmetric
random walk with point targets and an equal probability of moving
left with step size 2 and right with step size 1. $2+\mu_{num}$ is
calculated numerically at $t=10$ (blue dashes) and compared with
the asymptotic value of $2+\mu_{asym}$ (orange dots) and the corresponding
value of the CHSH parameter (green). The probabilities of arrival
form the same pattern as depicted in Figure \ref{fig:Four-targets},
those with darker shades take the range of values: $P_{\pm\pm}^{\hat{a},\hat{b}}=P_{\pm\pm}^{\hat{a}',\hat{b}}=P_{\pm\pm}^{\hat{a}',\hat{b}'}=P_{\pm\mp}^{\hat{a},\hat{b}'}=0.25-0.48$
whilst those with the lighter shades take the range of values: $P_{\pm\mp}^{\hat{a},\hat{b}}=P_{\pm\mp}^{\hat{a}',\hat{b}}=P_{\pm\mp}^{\hat{a}',\hat{b}'}=P_{\pm\pm}^{\hat{a},\hat{b}'}=0.02-0.25$.\label{fig:asm_random_walk}}
\end{figure}

Figure \ref{fig:asm_random_walk} depicts how the numerical $(2+\mu_{num})$
(blue dashes) and the analytical asymptotic value $(2+\mu_{asym})$
(orange dots) of the additional term, and the CHSH parameter (orange)
depend on $C(\hat{a},\hat{b})$ for the dynamics specified above.
The additional term appears to evolve in a similar manner to Figure
\ref{fig:sym_walk}. 

\begin{figure}[H]
\centering
\textbf{A)}\\

\includegraphics[width=1\columnwidth]{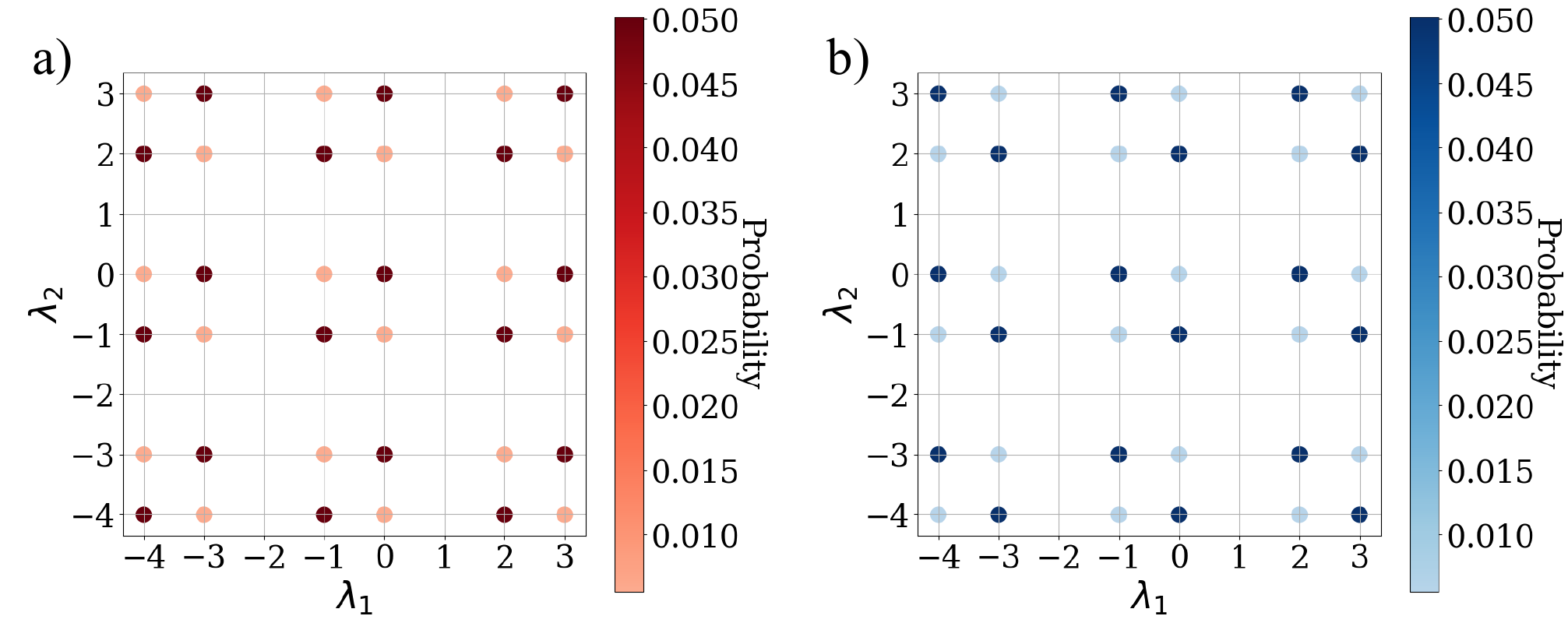}\\[1ex]

\textbf{B)}\\
\FloatBarrier
\includegraphics[width=1\columnwidth]{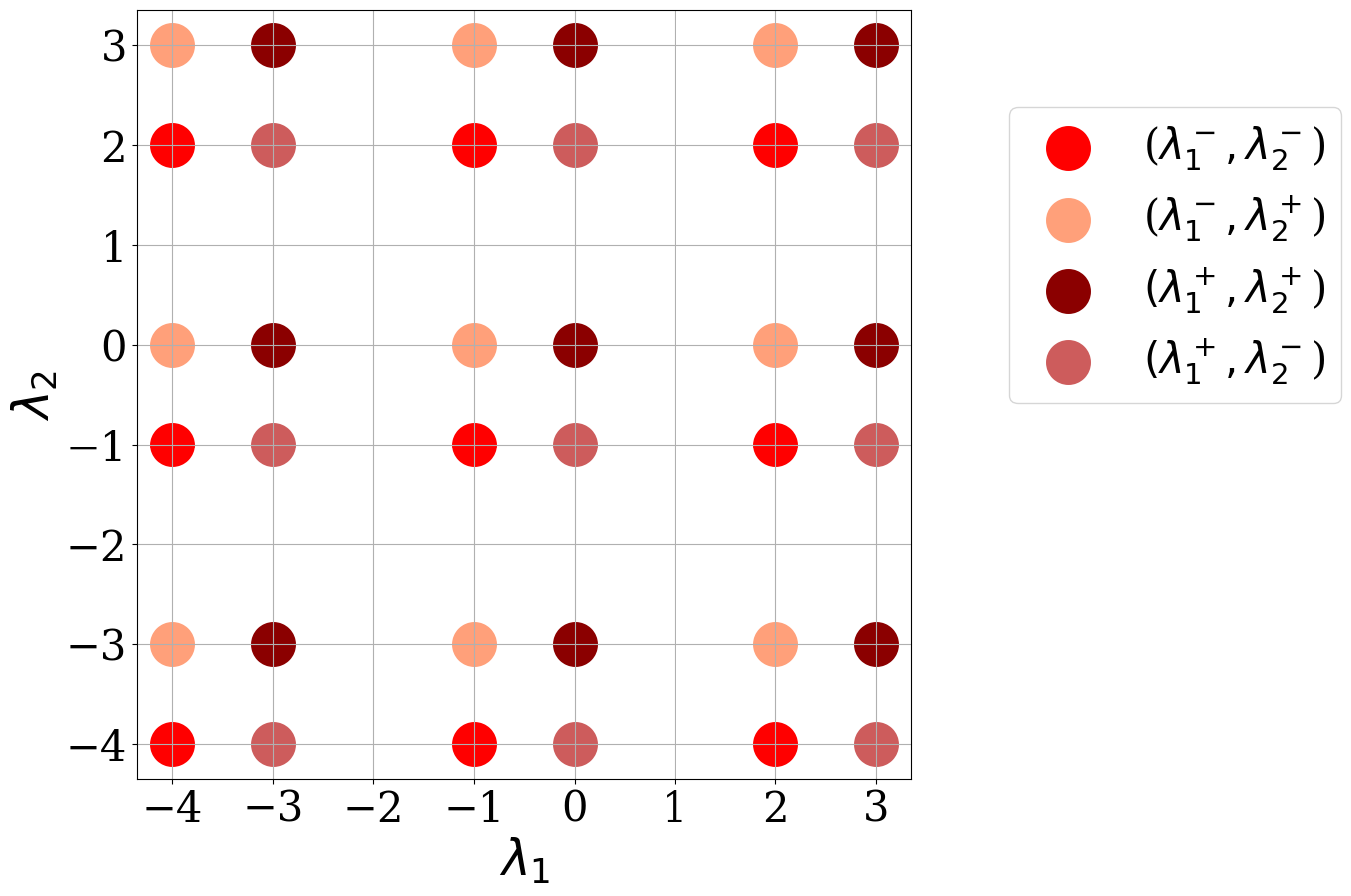}\\[1ex]

\caption{%
\textbf{A)} Probability distributions of the hidden variables $\lambda_{1,2}$
for measurement settings a) $(\hat{a},\hat{b})$ b) $(\hat{a}',\hat{b})$,
for an asymmetric random walk with step size 2 for moving left and
1 for moving right. Point targets were considered and an equal probability
of moving left or right. The post-measurement outcome probabilities
are $P_{\pm\pm}^{\hat{a},\hat{b}}=P_{\pm\pm}^{\hat{a}',\hat{b}}=P_{\pm\pm}^{\hat{a}',\hat{b}'}=P_{\pm\mp}^{\hat{a},\hat{b}'}=0.45$
and $P_{\pm\mp}^{\hat{a},\hat{b}}=P_{\pm\mp}^{\hat{a}',\hat{b}}=P_{\pm\mp}^{\hat{a}',\hat{b}'}=P_{\pm\pm}^{\hat{a},\hat{b}'}=0.05$,
calculated at $t=10$. 
\textbf{B)} Possible values of $\lambda_{1}$ and $\lambda_{2}$
leading to each of the four different targets $(\lambda_{1}^{\pm},\lambda_{2}^{\pm})$
for measurement setting $(\hat{a},\hat{b})$. The probabilities of
arrival are $P_{\pm\pm}^{\hat{a},\hat{b}}=P_{\pm\pm}^{\hat{a}',\hat{b}}=P_{\pm\pm}^{\hat{a}',\hat{b}'}=P_{\pm\mp}^{\hat{a},\hat{b}'}=0.45$
and $P_{\pm\mp}^{\hat{a},\hat{b}}=P_{\pm\mp}^{\hat{a}',\hat{b}}=P_{\pm\mp}^{\hat{a}',\hat{b}'}=P_{\pm\pm}^{\hat{a},\hat{b}'}=0.05$,
calculated at $t=10$. \label{fig:prob_dist_asym}}
\end{figure}

Figure \ref{fig:prob_dist_asym} (A) illustrates the probability distributions
of the hidden variable for measurement settings a) $(a,\hat{b})$
and b) $(\hat{a}',\hat{b})$. Similar distributions to a) are found
for $\rho(\lambda_{1},\lambda_{2}|\hat{a},\hat{b'})$ and $\rho(\lambda_{1},\lambda_{2}|\hat{a}',\hat{b'})$
. Similarly to the symmetric random walk case in Figure \ref{fig:prob_dist_sym}
(A), Figure \ref{fig:prob_dist_asym} (A) shows a checkerboard of
points of high and low probability. The values of $\lambda_{1}$ and
$\lambda_{2}$ with high and low probability in a) are opposite to
those in b), leading to a difference between $\rho(\lambda_{1},\lambda_{2}|\hat{a}',\hat{b})$
and the other three probability distributions. Note that in the symmetric
case, $\rho(\lambda_{1},\lambda_{2}|\hat{a},\hat{b}')$ had a different
pattern of probability to the other three distributions, whereas in
the asymmetric case, $\rho(\lambda_{1},\lambda_{2}|\hat{a}',\hat{b})$
is the probability distribution that is the odd one out, as a result
of the dynamics.

Unlike the symmetric case, there are values of $\lambda_{1}$ and
$\lambda_{2}$ that have a probability of adoption of zero across
all measurement situations at a given time-step and instead the values
of $\lambda_{1}$ and $\lambda_{2}$ with non-zero probability are
clustered in groups of four. Each possible value of $\lambda$ has
a cycle of three time-steps. For example, consider that at $t=1$,
1 is one of the many possible values of $\lambda$. At time-steps
$t=2$ and $3$, $\lambda=1$ is inaccessible and does not form one
of the possible values of $\lambda$, but at $t=4,8,16...$it will. 

Figure \ref{fig:prob_dist_asym} (B) depicts the different possible
values of $\lambda_{1}$ and $\lambda_{2}$ for measurement setting
$(\hat{a},\hat{b})$ leading to the four different targets. Similar
plots are formed for $\rho(\lambda_{1},\lambda_{2}|\hat{a}',\hat{b})$,
$\rho(\lambda_{1},\lambda_{2}|\hat{a},\hat{b'})$ and $\rho(\lambda_{1},\lambda_{2}|\hat{a}',\hat{b}')$.
Since for a given measurement situation, each target possesses a unique
set of $\lambda_{1}$ and $\lambda_{2}$ values, unique basins of
attraction must exist, leading to a significant value of the additional
term in Figure \ref{fig:asm_random_walk}. Note however, that unlike
the symmetric case, the basins of attraction leading to a target are
not defined by the parity of the target. In other words, there are
values of $\lambda_{1}$ and $\lambda_{2}$ of all parities that can
lead to an even-even target, and similarly for the other three targets.
Nonetheless, the values of $\lambda_{1}$ and $\lambda_{2}$ leading
to an even-even target, whilst not all even at an even time-step,
will be different to the values of $\lambda_{1}$ and $\lambda_{2}$
leading to the other three targets. The same can be said for the other
targets. There are therefore still two unique basins of attraction
for $\lambda_{1}$ and two for $\lambda_{2}$, leading to the targets
in a given measurement situation.

\FloatBarrier
\end{document}